\documentclass[%
 reprint,
superscriptaddress,
 amsmath,amssymb,
 aps,
]{revtex4-2}

\usepackage{graphicx}% Include figure files
\usepackage{dcolumn}% Align table columns on decimal point
\usepackage{bm}% bold math
\usepackage{qcircuit}
\usepackage{physics}
\usepackage{mathtools}
\usepackage{xcolor}
\usepackage{booktabs}
\usepackage{multirow}
\usepackage{tabularx}

\begin{document}

\preprint{APS/123-QED}

\title{Accelerating two-dimensional electronic spectroscopy simulations with a probe qubit protocol}% Force line breaks with \\

\author{José D. Guimarães}

\affiliation{%
Institute of Theoretical Physics and IQST, Ulm University, Albert-Einstein-Allee 11 89081, Ulm, Germany.
}%

\affiliation{%
Centro de Física das Universidades do Minho e do Porto, Braga 4710-057, Portugal
}%

\author{James Lim} 

\affiliation{%
 Institute of Theoretical Physics and IQST, Ulm University, Albert-Einstein-Allee 11 89081, Ulm, Germany.
}%
\author{Mikhail I. Vasilevskiy}

\affiliation{%
Centro de Física das Universidades do Minho e do Porto, Braga 4710-057, Portugal
}%

\affiliation{%
Intl. Iberian Nanotechnology Laboratory,
Av. Mestre Jos{\'e} Veiga s/n, Braga 4715-330, Portugal.
}%
\author{Susana F. Huelga}

\affiliation{%
Institute of Theoretical Physics and IQST, Ulm University, Albert-Einstein-Allee 11 89081, Ulm, Germany.
}%
\author{Martin B. Plenio}

\affiliation{%
Institute of Theoretical Physics and IQST, Ulm University, Albert-Einstein-Allee 11 89081, Ulm, Germany.
}%
\begin{abstract}
Two-dimensional electronic spectroscopy (2DES) is a powerful tool for exploring quantum effects in energy transport within photosynthetic systems and investigating novel material properties. However, simulating the dynamics of these experiments poses significant challenges for classical computers due to the large system sizes, long timescales and numerous experiment repetitions involved. This paper introduces the probe qubit protocol (PQP)-for quantum simulation of 2DES on quantum devices-addressing these challenges. The PQP offers several enhancements over standard methods, notably reducing computational resources, by requiring only a single-qubit measurement per circuit run and achieving Heisenberg scaling in detection frequency resolution, without the need to apply expensive controlled evolution operators in the quantum circuit. %Exclusively, this protocol enables the probing of dark-state dynamics within the quantum system during the detection stage of 2DES, a capability not found in standard protocols. 
The implementation of the PQP protocol requires only one additional ancilla qubit, the probe qubit, with one-to-all connectivity and two-qubit interactions between each system and probe qubits. We evaluate the computational resources necessary for this protocol in detail, demonstrating its function as a dynamic frequency-filtering method through numerical simulations. We find that simulations of the PQP on classical and quantum computers enable a reduction on the number of measurements, i.e. simulation runtime, and memory savings of several orders of magnitude relatively to standard quantum simulation protocols of 2DES. The paper discusses the applicability of the PQP on near-term quantum devices and highlights potential applications where this spectroscopy simulation protocol could provide significant speedups over standard approaches such as the quantum simulation of 2DES applied to the Fenna-Matthews-Olson (FMO) complex in green sulphur bacteria.

\end{abstract}

\maketitle

\section{Introduction}

Spectroscopy has long been a vital tool for probing photophysical processes in the quantum regime. Recent advancements in optical control \cite{cho2008coherent,hamm2011concepts, cong2020application, bustamante2021optical} have significantly enhanced our ability to examine quantum system properties with unprecedented accuracy. Two-dimensional electronic spectroscopy (2DES) \cite{hybl1998two, jonas2003two,brixner2005two, biswas2022coherent, fresch2023two} exemplifies this progress, enabling the detailed investigation of quantum effects of energy and charge transport in natural systems \cite{engel2007evidence, collini2010coherently, panitchayangkoon2010long, dostal2016situ,cao2020quantum}, as well as contributing to the development of new materials and optoelectronic devices \cite{scholes2006excitons,beljonne2009beyond,ostroverkhova2016organic,collini20212d}.

Simulating spectroscopy experiments is crucial for validating theoretical models, especially for complex systems with numerous degrees of freedom (DoF) that challenge classical simulation methods. These DoFs can originate from within the systems itself -- such as in light-harvesting complexes in photosynthetic systems \cite{maiuri2015ultra,ostroumov2013broadband}, which may consist of dozens to thousands of coupled molecules arranged in a variety of three-dimensional networks \cite{croce2018light,mohseni2014quantum} -- or from their environment, typically involving interactions with vibrational degrees of freedom \cite{prior2010efficient,christensson2012origin, chin2013role}. Classical numerical simulations \cite{ishizaki2009unified,prior2010efficient, de2015thermofield,rosenbach2016efficient,somoza2019dissipation,schulze2016multi,strathearn2018efficient} struggle to incorporate these correlations, often requiring simplifications that may compromise accuracy \cite{caycedo2022exact}.

Quantum processors offer a promising alternative \cite{altman2021quantum,kang2024seeking,guimaraes2024digital} that could potentially overcome the limitations faced by classical simulators. Despite the limitations of current noisy intermediate-scale quantum (NISQ) processors, recent progress indicates that near-term quantum computers may be able to simulate effectively multi-dimensional nuclear magnetic resonance spectroscopy experiments  \cite{li2017measuring,o2022quantum,seetharam2023digital,lamb2024ising,khedri2024impact}. This progress opens the door to the exploration of more complex settings, such as 2DES experiments in biomolecular complexes, including their their vibrational environments. The development of efficient quantum error correction protocols \cite{bravyi2024high, xu2024constant} and improved quantum hardware \cite{sivak2023real, bluvstein2024logical} further enhances the feasibility of such simulations.

\begin{figure}
\centering
\includegraphics[width=0.49\textwidth]{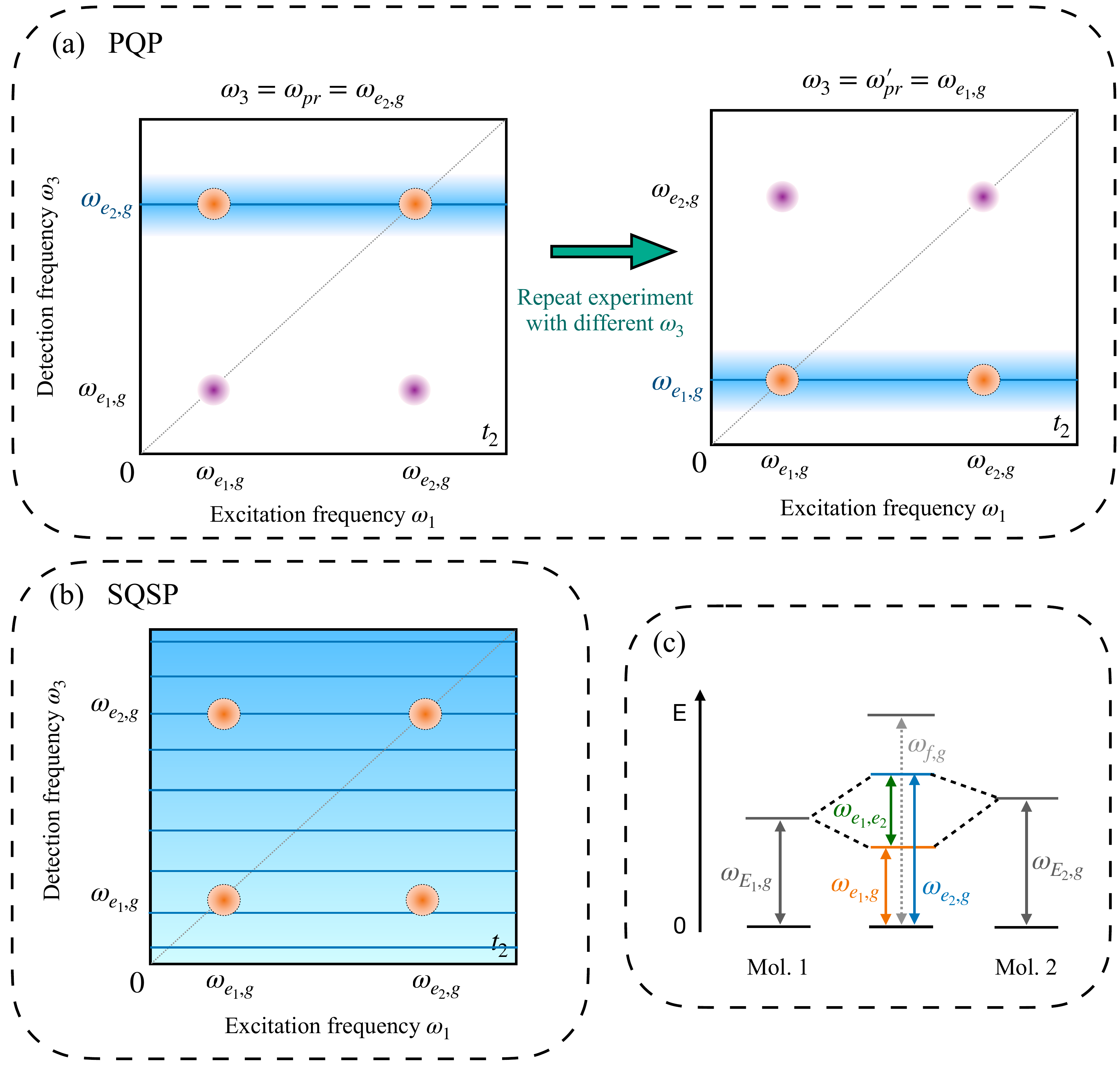}
    \caption{Schematic of a phase-cycled 2D spectra $\langle \hat{F}(\omega _1, t_2, \omega _3)\rangle$ for a specific $t_2$ using PQP (a) and the SQSP (b) for a molecular system with energy diagram (c). In the PQP simulation (a), we attach a probe qubit and repeat the experiment with different probe qubit energy gaps, $\omega_3=\omega_{pr}$ and $\omega_3=\omega'_{pr}$, in order to resolve the relevant peak amplitudes located at and close to the chosen $\omega_3$. This protocol functions as a dynamical frequency-filtering method. Orange (purple) circles represent resolved (unresolved) peak amplitudes. The width of blue lines corresponds to the frequency resolution $\Delta \omega_3$ of the peak amplitude estimation limited by the finite interaction time interval $t_3$ (more details in Sec.~\ref{sec: applicability}). (b) The SQSP resolves only the full 2D spectra with an uncertainty dictated by the number of data points taken over time interval $t_3$ (more details in Sec.~\ref{sec: standard}), without enabling the extraction of peak amplitudes in single, arbitrary detection frequencies $\omega_3$. The estimation of peak amplitudes over $\omega_1$ of both protocols follows the same procedure (with uncertainty limited by the number of data points taken over $t_1$). (c) Energy diagram of the molecular system considered in this work. Two molecules with excited states $\ket{E_1}$ and $\ket{E_2}$, respectively, are coupled via a electronic dipolar coupling leading to a four-level excitonic system as shown in the middle of the figure, with energy states $\ket{g}, \ket{e_1}, \ket{e_2}$ and $\ket{f}$, ordered respectively by ascending energies.}
    \label{fig:sqsp_pqp}
\end{figure}

In this work, we introduce the probe qubit protocol (PQP), a novel method for simulating 2DES experiments on quantum computers, such as noisy intermediate-scaling quantum (NISQ) devices and early-fault tolerant quantum computers (early-FTQC). Unlike the standard quantum simulation protocol (SQSP) \cite{lee2021simulation,huang2022variational,bruschi2024quantum,gallina2024simulating,kokcu2024linear,kharazi2024efficient}, the PQP bypasses the need to Fourier-transform the output 2DES signal over the detection stage by directly probing only a selected set of detection frequencies, as shown in Fig.~\ref{fig:sqsp_pqp}, and it does not require the implementation of expensive controlled evolution operators. Therefore, PQP offers significant advantages over the SQSP for 2DES, requiring only a single-qubit measurement per quantum circuit run and achieving Heisenberg-limited detection frequency resolution, surpassing the standard quantum limit (SQL) of the SQSP (see Table~\ref{tab:comparison}).
%Remarkably, the PQP allows the direct extraction of information about dark state dynamics—states with vanishing transition dipole moments—during the detection stage of a 2DES experiment.} 
The requirement of only one additional ancilla qubit in PQP, known as the probe qubit, with one-to-all connectivity to the system qubits, is particularly appealing for quantum computers with all-to-all qubit connectivity such as trapped ions or neutral atoms \cite{moses2023race,bluvstein2024logical,so2024trapped,sun2024quantum}. We summarize these improvements in Table~\ref{tab:comparison}.

The PQP builds on similar principles as other proposed near-term quantum phase estimation methods \cite{wang2012quantum,wang2016quantum,li2019quantum,stenger2022simulating,lu2021algorithms,somma2019quantum,lin2022heisenberg,dong2022ground,wan2022randomized,ding2023even}, but is specifically designed to extract coherence terms of the system density matrix in dynamical photophysical processes, as targeted by 2DES. Unlike the SQSP, which extracts the full 2D spectra (see Fig.~\ref{fig:sqsp_pqp}(b)) and suffers from a poor runtime scaling (as outlined in Table~\ref{tab:comparison}), PQP enables the extraction of peak amplitudes at selected detection frequencies as illustrated in Fig.~\ref{fig:sqsp_pqp}(a). This approach offers several advantages summarized in Table~\ref{tab:comparison} and includes the option for reconstruction of the full 2D spectra by iterating experiments across all detection frequencies.

The PQP is particularly advantageous in cases where the response of a quantum system are concentrated on a few detection frequency lines, as is often the case in studies of energy transport in photosynthetic systems \cite{engel2007evidence,collini2010coherently,panitchayangkoon2010long,dostal2016situ,cao2020quantum}. By focusing on a selected number of frequencies, the PQP can significantly accelerate the simulation process while maintaining high accuracy.
%Furthermore, the ability to probe dark state dynamics during 2DES experiments may open new avenues for applications, which we leave for future exploration. 

\renewcommand{\arraystretch}{1.7} % Increase row height by 50%
\begin{table*}
\centering
\begin{tabular}{|c|c|c|}
\hline
      &  \normalsize SQSP &  \normalsize PQP (this work) \\
     \hline
     \normalsize Number of qubits & \normalsize $n_{qub}$ & \normalsize $n_{qub}+1$ \\
    \hline
     \normalsize Number of single-qubit measurements & \normalsize $n_{qub}$ & \normalsize 1 \\
     \hline
     \normalsize Number of circuit executions & \normalsize $27 \times S_{SQSP} \times N_1 \times N_2 \times N_3$ & \normalsize $\approx 27 \times S_{PQP} \times N_1 \times N_2$ \\
     \hline
      \normalsize Frequency resolution $\Delta \omega_3$ & \normalsize $O\left(Q^{-1/2}\right)$ & \normalsize $O\left(Q^{-1}\right)$\\
     \hline
\end{tabular}
\caption{Comparison of the probe qubit protocol (PQP) introduced in this work and the standard quantum simulation protocol (SQSP) considered in this wrok. Parameter description: \emph{Number of qubits} and \emph{Number of single-qubit measurements} refer to the number of qubits and number of single-qubit measurements in each quantum circuit execution in the quantum device, respectively; \emph{Number of circuit executions} represents the number of circuits to be run in the quantum device in order to extract information about the peak amplitudes in a 2D spectra. $N_{j}$ is the number of desired data points to be collected in each simulated time interval $t_j$, with $j\in \{1,2,3\}$ ($j\in \{1,2\}$) in the SQSP (PQP) simulation and $S_{SQSP}\sim O\left(n_{qub}\varepsilon^{-2}\right)$ ($S_{PQP} \sim O(\varepsilon^{-2})$) is the required number of shots  to estimate each expectation value with observable measurement accuracy $\varepsilon$. \emph{Frequency resolution} $\Delta \omega_3$ denotes the scaling of the resolution of the detection frequency $\omega_3$ as a function of the number $Q$ of system Hamiltonian queries over the detection time interval $t_3$, i.e. the simulation runtime of the detection stage. The PQP simulation achieves Heisenberg-scaling whereas the SQSP only achieves the Standard Quantum Limit (SQL).}
\label{tab:comparison}
\end{table*}

The paper is organized as follows: Sec.~\ref{sec:intro_2DES} provides an overview of two-dimensional electronic spectroscopy (2DES) and the application of phase-cycling in conventional 2DES simulation protocols. Sec.~\ref{sec: standard} introduces the standard quantum simulation protocol (SQSP) for simulating phase-cycled 2DES on quantum computers, along with an estimation of the computational resources required for accurate results. Sec.~\ref{sec: pqp} presents the probe qubit protocol (PQP), detailing its implementation on quantum hardware, its advantages and disadvantages compared to other simulation protocols, and the conditions under which it should be applied. Sec.~\ref{sec: numerical} covers the numerical simulations of both PQP and SQSP, demonstrating similar quantitative results and suggesting that the PQP functions as a dynamic detection-frequency filter. The section also examines the effects of shot noise on the probe qubit within the PQP simulations. The estimation of the computational resources to perform a PQP simulation applied to the Fenna-Matthews-Olson (FMO) complex is also discussed. Finally, Sec.~\ref{sec:conclusion} concludes the paper by summarizing the findings and discussing potential future research directions.

\section{Introduction to 2DES}\label{sec:intro_2DES}
In a 2DES experiment setup, a quantum system is generally subjected to a sequence of three pulses, labelled by $j$, whose interaction with the sample is described by 
\begin{equation}
    \hat{H}_{I}^{(j)}(t) = -\sum_{m}\hat{\vec{\mu}}_{m} \cdot \vec{E}(t)\cos(\omega_p t - \vec{k}_{j} \cdot \vec{r}_{m}), \label{eq_HI}
\end{equation}
where $\omega_p$ is the frequency of the pulses and $\vec{E}(t)$ is the amplitude of the electric field of the pulse, a function of time (e.g. Gaussian), which we assume identical for all pulses $j$. $\vec{k}_{j}$ and $\vec{r}_{m}$ are the wavevector encoding the direction of the pulse $j \in \{1,2,3\}$ and the position vector of the quantum system $m$ relatively to a chosen reference frame, respectively \cite{mohseni2014quantum,mukamel1995principles}. For simplicity we assume that all pulses are identical, except for the direction of their wavevector.

%, each generating an interaction described by $\hat{H}_{p} = -\sum_{m}\hat{\vec{\mu}}_{m}\cdot \vec{E}$ between each site $m$ with transition dipole moment operator $\hat{\vec{\mu}}_{m}$ and each pulse with electric field $\vec{E}$ \cite{mohseni2014quantum}. Herein, we will assume, for simplicity, that all pulses are identical, except for their wavevector. Hence, we factorize the system-light interaction Hamiltonian in pulse-dependent and pulse-independent terms as follows,

In this work, we focus on the dynamics of the excitonic transport of photosynthetic systems and neglect doubly excited states on a single site. Hence we will consider the quantum system as a network of two-level molecules \cite{huelga2013vibrations,cao2020quantum}, i.e. qubits, and use the Hamiltonian \cite{caruso2009highly,rebentrost2009environment,huelga2013vibrations,zerah2021photosynthetic},
\begin{equation}
    \hat{H}_{S} = -\sum_{m} \frac{E^{(mol)}_{m}}{2} \hat{Z}_{m} + \sum_{n \neq m} \frac{J_{m,n}}{2} (\hat{X}_{m}\hat{X}_{n}+\hat{Y}_{m}\hat{Y}_{n}). \label{eq: HS}
\end{equation}
Here $\hat{X}, \hat{Y}$ and $\hat{Z}$ are Pauli operators, $E^{(mol)}_{m}$ is the energy of the excited state of the two-level molecule $m$ and $J_{m,n}$ is a dipolar electronic coupling. This Hamiltonian encodes the excitonic energy transport across a network of two-level molecules encoded in the qubits and it arises from excitonic creation and annihilation operators as shown in Eq.~\eqref{eq_HPR_exc}. Since we assume the system's sites to be two-level molecules (qubits), the transition dipole moment operator in Eq.~\eqref{eq_HI} can be defined as $\hat{\vec{\mu}}_{m}=\vec{\mu}_{m}\hat{X}_{m}$.

\subsection{The conventional protocol}\label{sec:2DES_conv_protocol}
The 2DES protocol proceeds as follows: the system, which we consider to be a pigment-protein complex, starts in the ground state, in equilibrium with its environment, and it is subjected to a sequence of three pulses at times $\tau_1,\tau_2$ and $\tau_3$. Since the intensity of pulses in 2DES experiments is typically low, the change of state of the system can be described in first order perturbation theory
according to $\hat{\rho}' \propto i\left[\sum_{m}\hat{\vec{\mu}}_{m},\hat{\rho}\right]$ following the Hamiltonian~\eqref{eq_HI}. After pulse $j$, the system evolves freely for a time $t_j=\tau_j-\tau_{j-1}$. 
%The 2DES protocol proceeds as follows: the system, which we consider to be a pigment-protein complex, starts in the ground state, in equilibrium with its environment, and the first pulse, at $t=0$, excites the system. \mbp{Since the intensity of pulses in 2DES experiments is typically low, the system-light interaction can be described in first-order perturbation theory of a dynamics governed by the Hamiltonian~\eqref{eq_HI}, leading to a density matrix after pulse $\hat{\rho}' \propto i\left[\sum_{m}\hat{\vec{\mu}}_{m},\hat{\rho}\right]$~\cite{mohseni2014quantum}, taking it to a superposition of the ground- and excited-states. The system is then left to evolve freely following the Hamiltonian $\hat{H}_{S}$ for a time $t_1=\tau_1$ in the time interval $[0,\tau_1]$. A second pulse then hits the sample at $\tau_1$ to convert coherence to populations and, subsequently, the system's density matrix evolves freely for a time $t_2 = \tau_2 - \tau_1$ in the time interval $[\tau_1,\tau_2]$ %from $t = \tau_1$ up to time $t = \tau_2$ for an additional time interval of $t_2 = \tau_2 - \tau_1$ following the Hamiltonian $\hat{H}_{S}$. The third pulse, at time $\tau_2$, then converts populations to coherence again which, subsequently, evolve freely for a time $t_3=\tau_3-\tau_2$ in the time interval $[\tau_2,\tau_3]$.% for a time interval $t_3=\tau_3-\tau_2$. 
Finally, the third-order polarization signal (written as a function of time intervals between pulses) $P(t_1,t_2,t_3) = \langle \hat{\vec{\mu}}(t_1,t_2,t_3)\rangle$ is measured (through optical heterodyne detection \cite{biswas2022coherent}), where $\hat{\vec{\mu}}=\sum_{m}\hat{\vec{\mu}}_{m}$ is the transition dipole moment of the pigment-protein complex and the angular brackets denote the quantum-mechanical expectation value. A 2D-Fourier transform is then applied to the time-domain signal in the variables $t_1$ and $t_3$ for fixed $t_2$, yielding two-dimensional plots of the polarization  $P(\omega_1,t_2,\omega_3)$. To accurately reproduce real-world experiments in our simulations, the polarization signal must be averaged over the static disorder present in the ensemble of bio-molecular complexes. Off-diagonal peaks in these spectra indicate energy transfer between different excitonic eigenstates, while the observation of oscillatory dynamics in specific off-diagonal peaks $\{(\omega _1,\omega _3)\}$ for varying $t_2$ delay times in the 2D spectra reveals information about the coherence of the system dynamics \cite{schlau2011ultrafast}. To analyze the system dynamics for specific third-order polarization signals, it is essential to eliminate the low-order signals as well as other undesired third-order signal contributions. Experimentally, this is typically achieved by placing the detector in some specific direction relative to the excitation pulses, thereby filtering out polarization signals that do not obey momentum conservation - commonly referred to as the phase-matching condition,
\begin{equation}
    \vec{k}_{s}= (-1)^{p_1}\vec{k}_{1} + (-1)^{p_2}\vec{k}_{2}+(-1)^{p_3}\vec{k}_{3}, \label{eq_phase_matching}
\end{equation}
which must be satisfied. Here $\vec{k}_{s}$ contains information about the direction at which the detector is placed and $p_{j}$ is either 0 or 1. Throughout this work, we will focus on the most common measured signals in experiments \cite{engel2007evidence, collini2010coherently, panitchayangkoon2010long}, namely, the rephasing signal $\vec{p}=(1,0,0)$ and the nonrephasing signal $\vec{p}=(0,1,0)$, where $\vec{p}=(p_1,p_2,p_3)$.

However, the conventional protocol to implement 2DES does not work well for small-size samples, e.g. a single bio-molecular complex which is modelled in our simulations, as it is the inverse of the size of the sample which dictates the momentum resolution \cite{mukamel1995principles}. Therefore, the simulation of small-size samples with the conventional protocol leads to the observation of an overlap of undesired polarization signals. This challenge is overcome through the use of \emph{phase-cycling} which, as we explain next allows us to isolate the response of a detector placed in a specific direction $\vec{k}_{s}$.

\subsection{Phase-cycling}\label{phase_cycling}
Here, we provide a short summary of the phase-cycling scheme applied to 2D spectroscopy. For a more comprehensive review, we refer the reader to Refs.~\cite{tan2008theory,pachon2015quantum,perdomo2012conformation}.

The phase-cycling scheme alleviates the issue of having unresolved polarization signals for small systems, by considering four pulses in a collinear geometry instead of three with different directions as done in the conventional 2DES protocol discussed above. The light-matter interaction Hamiltonian in a phase-cycled 2DES protocol is expressed as,
\begin{equation}
    \hat{H}_{I,pc}(t,\phi) = -\sum_{m}\hat{\vec{\mu}}_{m} \cdot \vec{E}(t)\cos(\omega_p t - \phi), \label{eq_HI_pc}
\end{equation}
where, in contrast to Eq.~\eqref{eq_HI}, each pulse $j \in \{1,2,3,4\}$ carries a relative phase $\phi_{j}$ instead of a phase defined by the pulses' wavevector. After applying the last pulse, $j=4$, the fluorescence of the sample is observed. As considered in previous works~\cite{perdomo2012conformation,pachon2015quantum}, we express the fluorescence observable as follows,
\begin{equation}
    \hat{F} = \Gamma_{1}\sum_{k_{1} \in \mathcal{M}_{1}} \ket{k_1}\bra{k_1} + \Gamma_{2} \sum_{k_{2}\in \mathcal{M}_{2}} \ket{k_2}\bra{k_2}+ \dots,\label{eq_F_obs}
\end{equation}
where $\ket{k_1}$ ($\ket{k_2}$) is a state in the single-excitation manifold $\mathcal{M}_{1}$ (double-excitation manifold $\mathcal{M}_{2}$) and higher-excitation manifolds may be also considered.~\cite{Note1} 
%$\Gamma_{l}$ is a parameter that can be optimized~\cite{perdomo2012conformation}, and defines how many photons are emitted via the relaxation process from the $l$-excitation manifold down to the ground-state manifold, i.e. $\Gamma_{l} \in [0,l]$. 
The physical meaning of the parameter $\Gamma_{1}$ is the quantum yield (QY), which is the relative probability of photon emission in the $e\rightarrow g$ transition. $\Gamma_{1}=1$ and $\Gamma_{2}=2$ implies a purely radiative exciton decay, an idealization that will be adopted here.
%For dark states (in the single-excitation manifold), $\Gamma_{1}$ is close to 0.
%Notice that the fluorescence observable  (\ref{eq_F_obs}) is just a sum of projection operators for all excited states.
%Physically, it means that all of them emit photons with the same QY.
The expectation value of the fluorescence observable in an experiment is obtained for several $t_j$, leading to the phase-dependent function $\langle\hat{F}(t_1,t_2,t_3,t_4,\vec{\phi}) \rangle$, dependent on the vector $\vec{\phi} = (\phi_{1},\phi_{2},\phi_{3},\phi_{4})$ spanning the phases of the four pulses.

In Eq.~\eqref{eq_HI_pc}, the phase $\phi$ can be seen as encoding the inner product of the pulse's wavevector and the position vector of the molecule with respect to some reference point, similar to Eq.~\eqref{eq_HI}. Since we intend to measure a particular type of 2D spectrum, e.g. the rephasing signal, in a phase-cycled experiment one must suitably choose the phases $\phi_{j}$ for all four pulses. In order to choose a particular signal to be observed through the use of phase-cycling, one runs several experiments cycling the phases around the circle for each pulse, such that the signal can be resolved \cite{tan2008theory}. In other terms, in each of the measurements for $t_1,\,t_2$ and $t_3$ (we assume $t_4=0$ without loss of generality), we need to cycle over a set of phases ${\phi}$ for each pulse and compute the following phase-independent expectation value function \cite{tan2008theory,pachon2015quantum}:
%\begin{align}
%    &\langle\hat{F}(t_1,t_2,t_3) \rangle = \label{eq_inv_Fourier_pc} \\
%    & \sum_{j_1,j_2,j_3,j_4}^{M} \langle\hat{F}(t_1,t_2,t_3,\phi_{j_1},\phi_{j_2},\phi_{j_3},\phi_{j_4})\rangle  \notag \\
%    &\times \exp{i(-1)^{(p_1+1)}\phi_{j_1}}\exp{i(-1)^{(p_2+1)}\phi_{j_2}} \notag\\
%    & \times
%\exp{i(-1)^{(p_3+1)}\phi_{j_3}}\exp{i(-1)^{(p_4+1)}\phi_{j_4}}\notag .
%\end{align}

\begin{align}
    \langle\hat{F}(t_1,t_2,t_3) \rangle &=\sum_{k=0}^{M^4} \langle \hat{F}(t_1,t_2,t_3;\vec{\phi}_{k})\rangle  \label{eq_inv_Fourier_pc} \\
    &\times \exp{-i\sum_{j}^{4}(-1)^{p^{(j)}_k}\phi_{k}^{(j)}} \notag .
\end{align}
where the first sum is over all permutations of a vector $\vec{\phi} \in \mathbb{R}^4$, where each element is a phase obtained from a set of size $M$. The total number of measurements for a data point ($t_1$, $t_2$, $t_3$) is $M^4$ assuming the same number $M$ of phases for each pulse ($j =1$, 2, 3 and 4). A particular signal is measured by appropriately choosing the phases $\vec{\phi}$ and the four-bit vector $\vec{p}$, emulating momentum conservation as in the conventional 2DES protocol in Eq.~\eqref{eq_phase_matching}, where we have, for instance,  $\vec{p}=(1,0,0,1)$ ($\vec{p}=(0,1,0,1)$) for the rephasing (non-rephasing) signal \cite{draeger2017rapid}. The inverse Fourier transform applied to $\langle\hat{F}(t_1,t_2,t_3;\vec{\phi}_{k})\rangle$ in Eq.~\eqref{eq_inv_Fourier_pc} enables us to resolve the four-bit vector $\hat{p}$ that encodes the desired polarization signal to be observed \cite{tan2008theory,pachon2015quantum}.
As shown in Refs.~\cite{tan2008theory,draeger2017rapid}, a pulse sequence $1\times 3\times 3\times 3$, where $\phi_{j} = \{0,2\pi/3,4\pi/3\}$ for $j=2$, 3 and 4 while $\phi_{1} = 0$, is the phase-cycling scheme with minimal number of measurements (i.e. 27) that can resolve the rephasing and non-rephasing signals (each can be obtained by merely changing the vector $\vec{p}$ in Eq.~\eqref{eq_inv_Fourier_pc} through post-processing of the measurement results). On the other hand, one may instead choose the cycle $3\times 3\times 3 \times 1$ with the same set of phases $\phi_{j} = \{0,2\pi/3,4\pi/3\}$, but now for $j=1$, 2 and 3 while $\phi_{4} = 0$, which also yields the rephasing and non-rephasing 2D spectra. After obtaining $\langle\hat{F}(t_1,t_2,t_3) \rangle$, a 2D-Fourier transform is applied over $t_1$ and $t_3$ to the measured signal, as in conventional 2D spectroscopy.

\section{Standard quantum simulation protocol of 2DES} \label{sec: standard}

In this section we describe the standard quantum simulation protocol (SQSP) of 2DES, defined as a protocol that requires the application of a 2D-Fourier transform over $t_1$ and $t_3$ to the simulated polarization signals \cite{lee2021simulation,huang2022variational,bruschi2024quantum,gallina2024simulating,kokcu2024linear}, in order to estimate its computational resource requirements. These requirements will later be compared to those needed for implementing the PQP protocol proposed in this work. This comparison allows us to identify concrete improvements offered by PQP over SQSP.

\subsection{Time-evolution simulation}

The SQSP of a 2DES experiment can be performed with several proposed time-evolution methods, namely via quantum signal processing and qubitization \cite{low2019hamiltonian,martyn2021grand}, holographic \cite{foss2021holographic,niu2022holographic}, variational methods \cite{lee2021simulation,huang2022variational,cirstoiu2020variational,mizuta2022local,berthusen2022quantum} or product formulas \cite{bruschi2024quantum,gallina2024simulating,kokcu2024linear,childs2021theory,csahinouglu2021hamiltonian}. In this work, we employ the Trotter-Suzuki product formula decomposition of the total evolution operator, which has received increasing interest in the past years as it promises significant resource reductions for higher-order approximations \cite{yang2022improved,morales2022greatly,ostmeyer2023optimised} and the proposal of randomized formulas \cite{nakaji2023qswift} which may be suitable for Hamiltonians with a large number of noncommuting terms. The conventional Trotter-Suzuki product formula describing the time evolution of a quantum system over time $t$ is formally expressed as
\begin{figure*}
\centering
\includegraphics[width=0.98\textwidth]{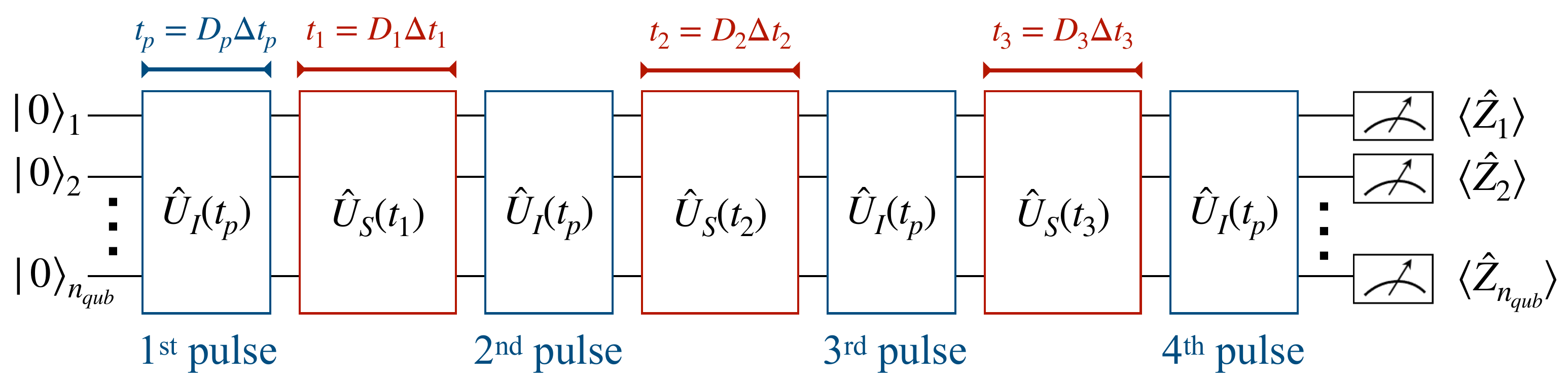}
    \caption{Overview of the implementation of the standard quantum simulation protocol of 2DES experiments. Qubits are initialized in the ground-state and evolved with two different types of evolution operators, namely $\hat{U}_{I}(t_p)$ encoding the system-pulse interaction and $\hat{U}_S(t_j)$ encoding the free system evolution, as defined by Eq.~\eqref{eq_Ui_trotter} and Eq.~\eqref{eq_Us_trotter}, respectively. Here, $t_p=D_p \Delta t_p$ and $t_j = D_j \Delta t_j$, with $j \in \{1,2,3\}$, and $D_p$ and $D_j$ denote the number of Trotter layers used for the respective evolution block. The Trotter time-steps $\Delta t_p$ and $\Delta t_j$ are also customizable for each evolution block. Lastly, the qubits are measured in the computational basis.}
    \label{fig:standard_simulation}
\end{figure*}
\begin{equation}
    e^{-i\hat{H}t} \approx \prod_{d=1}^{D} \hat{T}_{k}(\Delta t), \label{eq_trotter_equation_anyorder}
\end{equation}
where $\hat{T}_{k}(\Delta t)$ represents the unitary time evolution over a finite Trotter time-step $\Delta t$, described by the $k$-th order Trotter-Suzuki product formula, where the first- and second-order examples are given by 
\begin{align}
    \hat{T}_{1}(\Delta t)&=\prod_{j=1}^{N} e^{-i \hat{H}_{j}\Delta t},\label{trotter_equation}\\
    \hat{T}_{2}(\Delta t)&=\left[ \prod_{j=1}^{N} e^{-i \hat{H}_{j}\Delta t/2} \right]\left[\prod_{j'=N}^{1} e^{-i \hat{H}_{j'}\Delta t/2}\right].\label{eq_trotter_equation2}
\end{align}
The total time evolution is then described by $D = t/\Delta t$ Trotter layers.

\subsection{Implementation}\label{sec: impl_phase_cycling}
When simulating the phase-cycled 2D spectroscopy experiment, one must simulate two different types of evolution operators, one encoding the system-pulse interaction and another encoding the free evolution of the system. The free-evolution operator is expressed as,
\begin{equation}
        \hat{U}_{S}(t)=e^{-i \hat{H}_{S}t} \approx \prod_{d=1}^{D} \hat{T}^{(S)}_{k}(\Delta t),\label{eq_Us_trotter}
\end{equation}
with $\hat{H}_S$ given in Eq.~\eqref{eq: HS}. $\hat{U}_{S}(t)$ is encoded in quantum gates via (e.g. a second-order) Trotter-Suzuki product formula with Trotter layer $\hat{T}^{(S)}_{k}(\Delta t)$. In between pulses $j=1$, $2$ and $3$, the system  evolves freely in each of the time intervals interval $[\tau_{j-1},\tau_j]$ of duration $t_{j}$ and we may denote the respective evolution operator as $\hat{U}_{S}(t_{j})$. Therefore, for each time interval $t_j$, one may choose a different number $D_{j}$ of $k-$th order Trotter layers $\hat{T}_{k}^{(S)}(\Delta t_{j})$ with time step $\Delta t_j$.  The implementation scheme for a standard simulation of 2D spectroscopy with phase-cycling is shown in Fig.~\ref{fig:standard_simulation}.

The second type of evolution operator is defined according to the Hamiltonian $\hat{H}_{I} = \hat{H}_{S} + \hat{H}_{I,pc}(t_{p},\phi)$, which contains the light-matter interaction lasting for a time interval $t_{p}$ (assumed identical to all pulses) and may be expressed as,

\begin{equation}
        \hat{U}_{I}(t) = e^{-i \hat{H}_{I}t} \approx \prod_{d_p=1}^{D_p} \hat{T}_{k}^{(I)}(t,\Delta t_p),\label{eq_Ui_trotter}
\end{equation}
where, for instance, the first-order Trotter layer is defined as,
\begin{equation}
    \hat{T}_{1}^{(I)}(t,\Delta t_p) = e^{-i \hat{H}_{S}\Delta t_p}e^{-i \hat{H}_{I,pc}(t, \phi)\Delta t_p}. \label{eq_HIj_trotter_layer}
\end{equation}
The operator defined in Eq.~\eqref{eq_Ui_trotter} is applied for a total time $t_p$, which is the duration of each pulse acting on the system, defined by the amplitude of its electric field $\vec{E}(t)$) in Eq.~\eqref{eq_HI_pc}. We note that the Trotter-Suzuki product formula in Eq.~\eqref{eq_Ui_trotter} encodes the system-light interaction interleaved with the free-evolution of the system. Therefore, in Eq.~\eqref{eq_Ui_trotter}, the product formula provides a good approximation to the evolution operator when a sufficiently small Trotter time-step $\Delta t_p$ is chosen and a smooth Hamiltonian function $\hat{H}_{I,pc}(t_p, \phi)$ in Eq.~\eqref{eq_HI_pc} approximates the profile of the pulse. In other terms, one must satisfy the condition $\Delta t_p << 2\pi \max(\omega_p,\omega_{\rm max})^{-1}$, where $\omega_{\rm max}=\max_s(\Delta E_{s})$, being $\Delta E_{s}$ some energy gap of the system. Since the Hamiltonian $\hat{H}_{I,pc}(\Delta t_p, \phi)$ is composed of only single-qubit rotations around the $\hat{X}$ axis as defined by the transition dipole moment operator in Eq.~\eqref{eq_HI_pc}, the computational resources required to implement $\hat{U}_{I}(t_p)$ will be dominated by the Hamiltonian $\hat{H}_{S}$. On the other hand, in the approximate impulsive limit $t_p \to 0$, Eq.~\eqref{eq_Ui_trotter} is given only by a layer with $\hat{X}$ operators applied to all qubits without free-evolution.

\subsection{Resource estimation} \label{sec: res_estimation_sqsp}
For each time interval $t_1$, $t_2$, $t_3$ and each vector of angles $\vec{\phi}_{k}$, we measure the expectation value of the fluorescence observable defined in Eq.~\eqref{eq_F_obs}, which can be obtained by making single-qubit $\hat{Z}$ measurements on all system qubits. The total number of measurements is then defined as follows,
\begin{equation}
    M_{meas} = 27 \times S_{SQSP} \times N_1 \times N_2 \times N_3,\label{eq_meas_standard}
\end{equation} 
where $N_j$ is the number of sampled data points over each time interval $t_j$ and the constant $27$ comes from the number of measurements necessary to implement a phase-cycling scheme $3\times 3\times 3\times 1$. 

We denote $S_{SQSP}$ as the required number of shots to estimate the expectation value of the fluorescence observable on each data point $(t_1, t_2, t_3, \vec{\phi}_k)$ with SQSP. In what follows, we derive the scaling of $S_{SQSP}$ as a function of the system's size and observable measurement accuracy $\varepsilon$ (the observable here is all the $n_{qub}$ single-qubit $\hat{Z}$ operators as shown in Figure.~\ref{fig:standard_simulation}). We choose the measured observable $\hat{F}$ in Eq.~\eqref{eq_F_obs} such that it probes only population terms up to the double-excitation manifold as in typical simulations of 2DES applied to photosynthetic systems \cite{tan2008theory,pachon2015quantum,perdomo2012conformation} (i.e. we assume that above this manifold, populations are negligible). Hence the total number of states in the computational basis that can be measured is $|\mathcal{S}| = O(n_{qub}^2)$. Note that we must distinguish measurements of states in the single- and double-excitation manifolds because of the different choices of coefficients $\Gamma_{l}$ with $l=1,2$ in Eq.~\eqref{eq_F_obs}. The probability to measure a state in the single-excitation (double-excitation) manifold via post-selection of the computational basis measurement is $p_{\mathcal{M}_{1}} = n_{qub}/|\mathcal{S}| = O(n_{qub}^{-1})$ ($p_{\mathcal{M}_{2}} = O(n_{qub}^2/|\mathcal{S}|) = O(1)$). Therefore, to estimate the expectation value estimation of the fluorescence observable in Eq.~\eqref{eq_F_obs} for a desirable observable measurement accuracy $\varepsilon$, we require a number of shots given as follows \cite{cai2023quantum},
\begin{equation}
    S_{SQSP} \sim \varepsilon^{-2}(p_{\mathcal{M}_{1}}^{-1}+p_{\mathcal{M}_{2}}^{-1}) = O\left(\frac{n_{qub}}{\varepsilon^2}\right). \label{eq_S_sqsp}
\end{equation}

Let us consider now the scaling of each number $N_{j}$ of samples for $j \in \{1,3\}$ as a function of the desired frequency resolution on the 2D spectra. Given some desirable frequency resolution $\Delta \omega_j$ for $j \in \{1,3\}$ and maximum frequency $\omega_j^{(max)}$ to be observed on the 2D spectra, we have $\Delta \omega_j = 2\pi / (N_j \Delta t_j)$ and  $\omega_j^{(max)} = \pi /\Delta t_j$, where $\Delta t_j$ is the time interval between samples during $t_j$. Hence, to obtain a frequency resolution $\Delta \omega_j$ and maximum frequency $\omega_j^{(max)}$ on the $\omega_j$ axis of the 2D spectra with $j \in \{1,3\}$, one must collect a number of data points as follows,
\begin{equation}
    N_{j}= \frac{2\omega_j^{(max)}}{\Delta \omega_j}.\label{eq_Nj_standard}
\end{equation}
%For instance, in typical photosynthetic systems, one may need $\omega_j^{(max)} \sim 10^5$~$cm^{-1}$ and $\Delta \omega_j \sim 10^1$~$cm^{-1}$, leading to $M_{meas} \sim 10^9$ measurements, with $N_2 \lesssim 10^1$.

Let us look now to the quantum computational resources required to implement the SQSP. In the quantum circuit, one needs $n_{qub}$ qubits, i.e. the total number of two-level molecules, initialized at $\ket{0}$, and a maximum circuit depth of $D_{max} = D_{1}+D_{2}+D_{3}+4D_{p}$, where $D_{p}$ ($D_j$) is the number of Trotter layers of $\hat{U}_{I}(t_p)$ ($\hat{U}_{S}(t_j)$). The circuit depth can also be given as a function of time $t_j$, Trotter approximation error $\epsilon_{trot}$ and the $k_j$th order of the product formula \cite{childs2021theory} as $D_{j} = O(t_j^{1+1/k_j}/\epsilon_{trot}^{1/k_j})$. Since $t_p$ has constant circuit depth $D_{p}$ relatively to frequency resolution $\Delta \omega_{j}$ and $t_{j} = 2\pi/\Delta \omega_{j}$ for $j \in \{1,3\}$, the maximum circuit depth for the simulation, can be expressed as follows,
\begin{equation}
         D_{max} =  O  \left(\sum_{j=1,3}\frac{1}{\Delta \omega_{j}(\Delta \omega_{j}\epsilon_{\rm trot})^{1/k_j}} + \frac{t_2^{1+1/k_2}}{\epsilon_{trot}^{1/k_2}}\right). \label{eq: D_max_sqsp}
\end{equation}
This equation shows that in the limit of high-order Trotter product formulas, the leading term for the maximum circuit depth is $O(n_{qub}[1/\Delta \omega_1+1/\Delta \omega_3+t_2])$, i.e. it is inversely proportional to the frequency resolution of $\omega_1$ and $\omega_3$.

Lastly, we show that this approach leads to the standard quantum limit (SQL) when applied to the estimation of peak amplitudes on the full 2D spectra. The number of required queries to the system Hamiltonian $\hat{H}_{S}$ in $\hat{U}_{S}(t_j)$ for $j\in \{1,3\}$ is proportional to $Q_S = \sum_{q=1}^{N_{j}} q = N_{j}(N_{j}+1)/2$. Since $N_{j} = O(D_j)$, we have that, for $j=3$, $D_3 = O(\Delta \omega_{3}^{-1}[\Delta \omega_{3}\epsilon_{\rm trot}]^{-1/k_3})$. For product formulas with sufficiently high order in $k_3$ and low Trotter error $\epsilon_{\rm trot}$, we find that the scaling of the precision $\Delta \omega_3$ as a function of the number of queries to the Hamiltonian leads to the SQL, namely,
\begin{equation}
    \Delta \omega_{3} = O\left(\frac{1}{\sqrt{Q_{S}}}\right).
\end{equation}

\begin{figure*}
\centering
\includegraphics[width=0.98\textwidth]{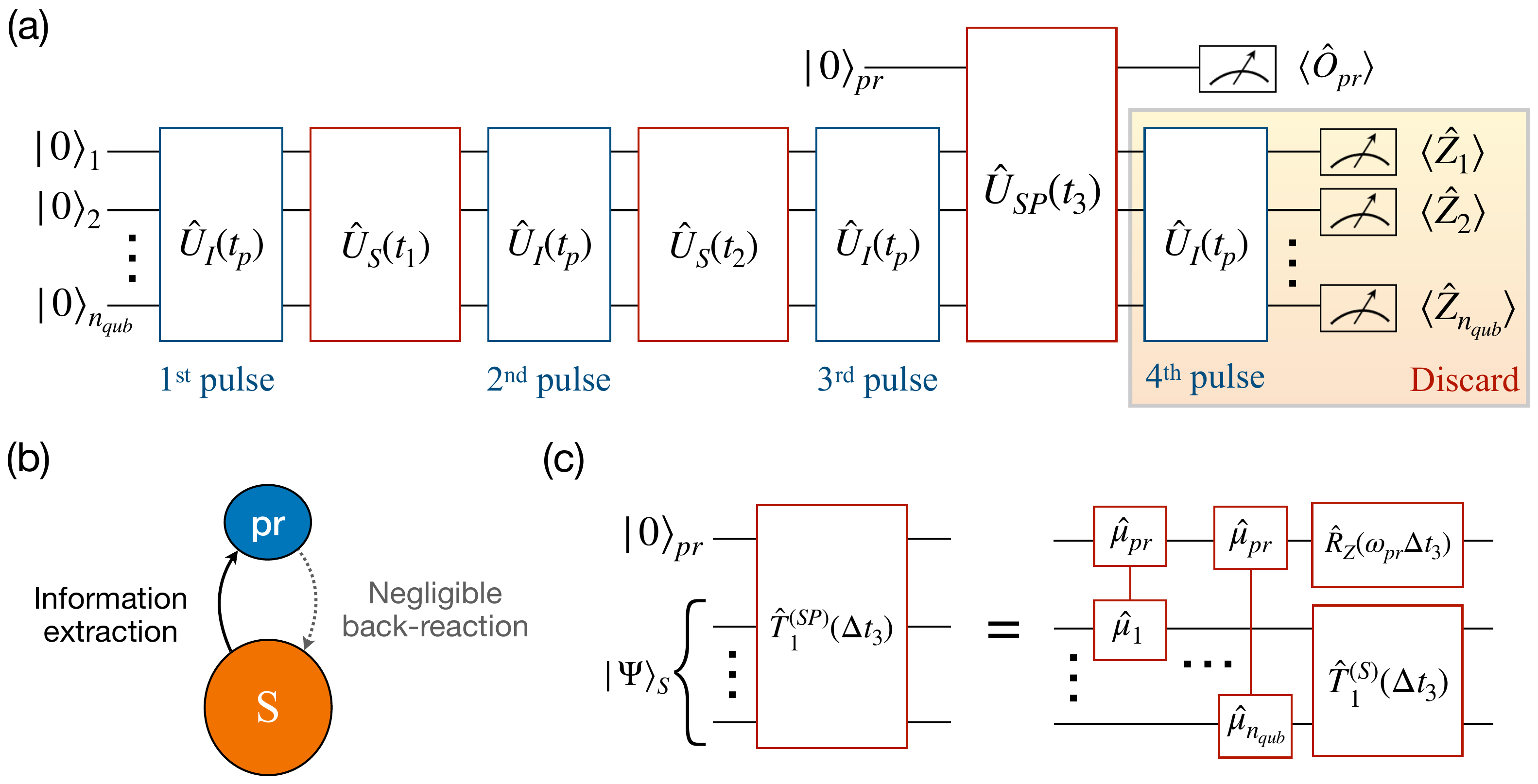}
    \caption{Overview of the probe qubit protocol to simulate 2D spectroscopy experiments on a quantum computer. (a) Quantum circuit with the probe qubit interacting with the system during the time interval $t_3$. The operations in the orange box do not need to be applied in this protocol, hence they can be discarded. The Pauli observable $\hat{O}_{pr}$ is measured after $t_3$. (b) The system-probe interaction consists of a weak energy transfer process. Since the probe qubit is initialized in the ground-state $\ket{0}_{pr}$, one can control the back-reaction on the system from the probe to the system so that it remains negligible over the time interval $t_3$ unless $t_3$ is too long (see Appendix A). (c) Trotter layer used to implement the evolution over $t_3$ with system-probe interactions. The probe weakly couples to each molecule via an electronic dipole interaction during $t_3 = D'_3 \Delta t_{3}$, where $D'_3$ is the number of Trotter layers over $t_3$ in this scheme and $\Delta t_{3}$ is the Trotter time-step. $\ket{\Psi}_{S}$ denotes the system's state after the third pulse interaction. The rotation applied to the probe qubit is defined as $\hat{R}_{Z}(\theta)=e^{i\theta\hat{Z}/2}$.}
    \label{fig:probe_scheme}
\end{figure*}

\section{The probe qubit protocol}\label{sec: pqp}
In view of achieving a more efficient quantum simulation approach than the SQSP presented above, in this section we propose a new protocol, where we add an ancilla qubit during the time interval $t_3$ that weakly interacts with the system and extracts some spectral information from it as diagrammatically illustrated in Fig.~\ref{fig:probe_scheme}(b). We call this method, the \emph{probe qubit protocol} (PQP).

\subsection{Implementation}
The probe qubit protocol (PQP) is illustrated in Fig.~\ref{fig:probe_scheme}(a). We perform the standard simulation as done in the SQSP until time $t=t_1+t_2+3t_p$, i.e. until the third pulse finishes its interaction with the system. We then initialize an ancilla qubit, which we call \emph{probe qubit}, in the $\ket{0}_{pr}$ state, and we let it interact weakly with the system over a time $t_3$. Information about the spectral features of the system will be extracted to the probe qubit over $t_3$, and then we measure a Pauli observable $\hat{O}_{pr}=\hat{X}_{pr}$ or $\hat{O}_{pr}=\hat{Y}_{pr}$. The system-probe qubit interaction is expressed as,

\begin{equation}
    \hat{H}_{PR} = -\frac{\omega_{pr}}{2}\hat{Z}_{pr}+\sum_{m} {\frac{J_{m}^{(pr)}}{2}}(\hat{X}_{pr}\hat{X}_{m}+\hat{Y}_{pr}\hat{Y}_{m}).\label{eq: HPR}
\end{equation}

Here, $\omega_{pr}$ is the energy gap of the probe qubit and $J_{m}^{(pr)}$ is the interaction strength between the probe and the qubit $m$ of the system.
The system-probe interaction $\hat{X}_{pr}\hat{X}_{m}+\hat{Y}_{pr}\hat{Y}_{m}$ encodes energy transport between each molecule $m$ and the probe. It can be interpreted as an electronic dipolar coupling between each molecule in the system and probe, with $\hat{\mu}_{pr}$ and $\hat{\mu}_{m}$ denoting the transition dipole moments of the probe and molecule $m$, respectively. This interaction is illustrated in Fig.~\ref{fig:probe_scheme}(c). Therefore, the evolution of the system and probe during $t_3$ is given by the Hamiltonian $\hat{H}_{SP} = \hat{H}_{PR}+\hat{H}_{S}$, which is implemented in a quantum circuit with a Trotter-Suzuki product formula as follows,
\begin{equation}
        \hat{U}_{SP}(t_3)=e^{-i \hat{H}_{SP}t_3}\approx \prod_{d_3=1}^{D'_3} \hat{T}^{(SP)}_{k}(\Delta t_3),\label{eq_Usp_trotter}
\end{equation}
where we have a number $D'_{3}$ of Trotter layers and Trotter time-step $\Delta t_3$. In Fig.~\ref{fig:probe_scheme}(c), we illustrate how a Trotter layer $\hat{T}^{(SP)}_{k}(\Delta t_3)$ with $k=1$ can be implemented. In Appendix~\ref{sec: probe_system_dynamics}, we calculate a perturbative expansion of the evolution operator shown in Eq.~\eqref{eq_Usp_trotter}, and find that the probe introduces energy exchanges between system's energy eigenstates in different excitation manifolds with energy gaps close to $\omega_{pr}$. 

We implement the probe qubit protocol using the $3\times 3\times 3\times 1$ phase-cycling scheme introduced in Sec.~\ref{phase_cycling}, where the first three pulses cycle over the phases $\{0,2\pi/3,4\pi/3\}$. With this scheme, we can select a desired type of spectrum (e.g. rephasing) to be observed by controlling the first three pulses' phases and discarding the last pulse (containing a constant phase across experiments). This is illustrated in Fig.~\ref{fig:probe_scheme}(a). After obtaining the expectation values $\langle \hat{O}_{pr} \rangle$ for data points given by the parameters $(t_1,t_2,\vec{\phi}_{k})$, we perform the inverse Fourier transform over the phases $\vec{\phi}_{k}$ as discussed in Sec.~\ref{phase_cycling}, but with $\hat{O}_{pr}$ as the measured Pauli observable and constant time $t_3$. 

\subsection{Information extraction}\label{sec: info_extraction_PQP}
Let us now analyze the type of information that we extract by measuring the probe qubit. We assume that we have a noiseless probe channel and the initial system's density matrix is (at $t_3 = 0$) given by $\hat{\rho}_{SP}(t_1, t_2) = \sum_{l,l'} \beta_{l,l'}(t_1, t_2)\ket{E_{l}}\bra{E_{l'}}_{S}\otimes \ket{0}\bra{0}_{pr}$ with $\ket{E_{l}}$ denoting the system states in the energy basis. As shown in the Appendix~\ref{sec: coherence_term}, the measurement of the observable $\hat{O}_{pr}=\hat{Y}_{pr}$ at a sufficiently long time $t_3$, in the weak system-probe interaction regime yields,
\begin{equation}
    \langle\hat{Y}_{pr}(t_1,t_2,\omega_{pr})\rangle_{t_3} \approx 2t_3\sum_{\omega_{l,l'}=\omega_{pr}} J_{l',l}^{(pr)}\Re[\beta_{l',l}(t_1,t_2)]\,,\label{eq_evol}
\end{equation}
where $\omega_{l,l'}=E_{l}-E_{l'}$ is an energy gap of the system and $J_{l',l}^{(pr)}= \sum_{m}J_{m}^{(pr)}\bra{E_{l'}}[(\hat{X}_{m}+i\hat{Y}_{m})/2]\ket{E_{l}}$ is the effective probe-system coupling strength, defined in the system's energy eigenbasis, which generates a system's transition from an eigenstate $\ket{E_{l}}$ to a state $\ket{E_{l'}}$ in one of 
the adjacent excitation manifolds (see Appendix \ref{sec: coherence_term} for more details). Here $\beta_{l',l}(t_1,t_2)$ is the coefficient of the coherence term at $t_3=0$ that we want to know. The sum in Eq.~\eqref{eq_evol} runs over all the coefficients of the system's coherence terms, $\beta_{l,l'}(t_1, t_2)\ket{E_{l}}\bra{E_{l'}}_{S}$, weighted by system-probe coupling, $J_{l',l}^{(pr)}$, taking into account all possible transitions in the system with the energy $\omega_3=\omega_{l,l'}=\omega_{pr}$. Therefore, information on the real ($\hat{O}_{pr} = \hat{Y}_{pr}$) or imaginary ($\hat{O}_{pr} = \hat{X}_{pr}$) component of each coherence term evolving with the frequency $\omega_{pr}$ is transferred to the probe via a (system-probe) coupling-weighted average. One may choose all $J_{m}^{(pr)}$ to be identical, leading to the typical situation of a phase-cycled 2DES experiment. On the other hand, choosing different $J_{m}^{(pr)}$ may, for instance, enable the observation of dynamics of excitonic dark states, i.e. system energy states with vanishing transition dipole moment.

The difference on the information obtained from a probe-qubit protocol and the SQSP presented in Sec.~\ref{sec: standard}, is that PQP obtains only partial information on the 2D spectra, whereas SQSP obtains the full 2D spectra. Fig.~\ref{fig:sqsp_pqp} illustrates the action of PQP as an effective dynamic frequency-filtering method over detection frequencies $\omega_3$. The use of the probe qubit permits to avoid Fourier-transforming over $t_3$ by probing only a limited (desired) set of frequencies $\omega _3$.
 
 To be more precise, let us discuss the full simulation workflow using the PQP, which involves executing several experiments with different probe energies $\{\omega_{pr}\}$ and extracting the real and imaginary parts of peak amplitudes for an arbitrary set of detection frequency lines $\{\omega_{3}=\omega_{pr}\}$ in the 2D spectra. This is carried out by obtaining the expectation values $\langle \hat{Y}_{pr}(t_1,t_2,\omega_3,\vec\phi_{j})\rangle_{t_3}$ and $\langle \hat{X}_{pr}(t_1,t_2,\omega_3,\vec\phi_{j})\rangle_{t_3}$ according to the phase-cycling technique, followed by applying an inverse Fourier transform over the phases $\vec\phi_{j}$ for each expectation value, leading to phase-independent $\langle \hat{Y}_{pr}(t_1,t_2,\omega_3)\rangle_{t_3}$ and $\langle \hat{X}_{pr}(t_1,t_2,\omega_3)\rangle_{t_3}$. After collecting sufficient data points over $t_1$ and $t_2$, one performs a Fourier transform over $t_1$ to the expectation values, leading to $\langle \hat{Y}_{pr}(\omega_1,t_2,\omega_3)\rangle_{t_3}$ and $\langle\hat{X}_{pr}(\omega_1,t_2,\omega_3)\rangle_{t_3}$. To obtain the same information as in the SQSP, but restricted to the detection frequency lines $\{\omega_{3}=\omega_{pr}\}$ in the 2D spectra, we sum the Fourier-transformed expectation values $\langle \hat{Y}_{pr}(\omega_1,t_2,\omega_3)\rangle_{t_3}-i\langle\hat{X}_{pr}(\omega_1,t_2,\omega_3)\rangle_{t_3}$.

 \subsection{Conditions of applicability}\label{sec: applicability}
As mentioned above, the bounds on the time interval of duration $t_3$ for which the system and probe interact must be defined in order to extract information from the system with a negligible back-reaction introduced by the probe. In Appendix~\ref{sec: probe_system_dynamics}, we show that the system-probe interaction time $t_3$ must be bounded as follows,
\begin{equation}
    O\left(2\pi \omega_{pr,s,s'}^{-1}\right) < t_3 < O\left(\left[n_{qub}\max_{m} [J_{m}^{(pr)}]\right]^{-1}\right). \label{eq_condition_t3_main_text}
\end{equation}
Here $\omega_{pr,s,s'}=\min_{s,s'}(|\omega_{pr}-\omega_{s,s'}|)$, where $\omega_{s,s'}=E_s-E_{s'}$ is the energy gap of the system closest to the one we are trying to probe, i.e. $\omega_{pr}$. The upper bound in Eq.~\eqref{eq_condition_t3_main_text} comes from the perturbative approach of the system-probe interaction, which imposes a worst-case limit on how large $t_3$ can be in order to reliably extract information from the system without non-negligible back-reaction effects. Choosing $\max_{m}[J_{m}^{(pr)}]\leq O(n_{qub}^{-1})$ enables one to achieve a constant upper bound on $t_3$. In view of probing a particular non-degenerate energy transition $\omega_3 \neq \omega_{s,s'}$ in a 2D spectra, the condition in Eq.~\eqref{eq_condition_t3_main_text} is always possible to achieve by appropriately lowering the system-probe couplings $J_{m}^{(pr)}$.

So far, we have considered a noiseless probe qubit. However, in realistic implementations of the protocol, noise will be introduced. In this case, the probe qubit will suffer decoherence for the entire duration of mutual interaction, which will affect the measurement outcome of $\langle\hat{Y}_{pr}\rangle$ and $\langle\hat{X}_{pr}\rangle$. As it is common for phase estimation methods \cite{lin2022heisenberg,ding2023even,dong2022ground}, noise is damaging to the protocol because the system-probe information extraction process relies on coherence retention on the probe and its static initial state to be $\ket{0}_{pr}$. Therefore, amplitude and phase damping relaxation noises are specially harmful to this protocol, leading to the loss of correlations between the system and the probe along $t_3$, which will reduce the observable measurement outcomes $\langle\hat{Y}_{pr}\rangle$ and $\langle\hat{X}_{pr}\rangle$ with increasing $t_3$. In such cases, choosing a shorter $t_3$ may be more appropriate at the cost of algorithmic errors (introduced by not satisfying the lower bound condition in Eq.~\eqref{eq_condition_t3_main_text}).

\subsection{Implementation on near-term quantum computers}
The quantum simulation with the PQP can be applied in the framework of currently available quantum hardware technologies. In platforms with nearest-neighbor qubit connectivity as, for instance, in superconducting quantum devices, one may implement SWAP gates to implement the system-probe interaction evolution over the time interval $t_3$. However, such simulation implementation is expected to contain higher degrees of noise than in platforms with one-to-all qubit connectivity due to the inclusion of SWAP operators in each Trotter layer. On the other hand, trapped-ion or neutral-atom platforms are particularly suited for this method due to the possibility of having all-to-all qubit connectivity \cite{pogorelov2021compact,moses2023race,bluvstein2024logical}. Additionally, in our proposed protocol, we require only a single-qubit measurement at the end of the circuit, which may be executed in trapped-ion or neutral-atom technologies with low error rates by moving a single qubit at the end of the circuit to a readout zone \cite{pogorelov2021compact,moses2023race,bluvstein2024logical}. The presence of native vibrational modes spanning the array of qubits and its control in trapped-ion quantum devices may also enable the simulation of excitonic energy transport in photosynthetic systems in the presence of vibrational mode interactions as demonstrated in Ref.~\cite{lemmer2018trapped,gorman2018engineering,whitlow2023quantum}. Such simulation is known to be extremely challenging for classical computers \cite{caycedo2022exact, lorenzoni2024systematic, chin2013role}. One may also use quantum algorithms such as Q-TEDOPA \cite{guimaraes2024digital} that enable the simulation of non-perturbative dynamics of open systems, where vibrational mode interactions can be implemented, for instance, by the method recently proposed in Ref.~\cite{katz2023programmable}. Analog quantum simulation techniques \cite{lemmer2018trapped,macdonell2021analog} may also be employed together with our proposed PQP approach to simulate vibronic dynamics in photosynthetic systems \cite{caycedo2022exact} or to perform quantum process tomography \cite{pachon2015quantum}.

\subsection{Resource estimation}\label{sec:res_est_pqp}

We conclude the description of the PQP by showcasing its advantages and disadvantages in comparison with the SQSP presented in Sec.~\ref{sec: res_estimation_sqsp}. We start by estimating the required quantum computational resources to obtain the peak amplitudes on a detection frequency line in a 2D spectra with PQP. Relatively to the SQSP, the number of qubits in this protocol is increased from $n_{qub}$ to $n_{qub}+1$. On the other hand, the maximum circuit depth is slightly decreased to $D_{max} = D_1 + D_2 + D'_3 + 3D_p $ due to the discard of the last system-pulse interaction layer as shown in Fig.~\ref{fig:probe_scheme}(c). Note that, in general, the circuit depth $D'_3$ over $t_3$ in the PQP will be different than the one estimated for the SQSP in Sec.~\ref{sec: res_estimation_sqsp} by a constant factor. Let us assume one-to-all qubit connectivity for the probe-system interaction (meaning that the probe can directly interact with all system's qubits) is possible with constant overhead as approximately attained in trapped-ion and neutral atom platforms \cite{bluvstein2024logical,moses2023race}. We consider a $k_j $th-order Trotter-Suzuki product formula during the time interval $t_j$. Per the lower bound in Eq.~\eqref{eq_condition_t3_main_text}, the maximum allowed detection frequency resolution is $\Delta \omega_3 \approx \omega_{pr,s,s'} = \min(|\omega_{pr}-\omega_{s,s'}|)$, i.e. dependent on the closest system energy gap $\omega_{s,s'}$ to $\omega_{pr}$. However, in general, we have that $\Delta \omega_3 = O(t_3^{-1})$ in the PQP and choosing a longer time $t_3$ (i.e. $D'_3$) leads to the reduction of $\Delta \omega_3$. In these conditions, the maximum circuit depth achieved by the PQP as a function of frequency resolutions $\Delta \omega_1$ and $\Delta \omega_3$ in the asymptotic regime is the same as the one obtained for the SQSP in Eq.~\eqref{eq: D_max_sqsp}. However, due to hidden constants in the big-O notation of the lower bound in Eq.~\eqref{eq_condition_t3_main_text}, we find that, in practice, a constant multiplicative factor $\alpha>1$ will lead to $D'_3 > D_3$, meaning that the circuit depth for the PQP will be generally higher than that of the SQSP.

Regarding the estimation of the number of measurements, the probe qubit is the only qubit that needs to be measured in the PQP. Therefore, the number of required single-qubit measurements in one circuit execution is just one, in comparison to the SQSP which scales linearly with system's size. On the other hand, the major improvement of the PQP over the SQSP arises in the total number of measurements required to estimate peak amplitudes of the 2D spectra. In this protocol, one needs a number of measurements given as,
\begin{equation}
    M_{meas}^{(pr)}=27 \times S_{PQP} \times N_1 \times N_2 \times N_{freq}, \label{eq_Mmeas_pr}
\end{equation}
to estimate the real or imaginary part of a coherence term coefficient of the system's density matrix, where the desired number of detection frequency lines to extract peak amplitudes in the 2D spectra is given by $N_{freq}$. This result should be contrasted to the one obtained with SQSP in Eq.~\eqref{eq_meas_standard}, where a frequency resolution dependency on $N_3$ exists.

We denote $S_{PQP}$ as the number of shots to estimate the real or imaginary part of a coherence term $\beta_{l,l'}(t_1,t_2)$ for a given constant $t_3$ with (single-qubit) observable measurement accuracy $\varepsilon$, where $\omega_{l,l'} = \omega_{pr}$ is the probe's energy gap (see Eq.~\eqref{eq_evol}). In what follows, we derive the scaling of $S_{PQP}$ as a function of system's size and $\varepsilon$. To calculate the variance of $\beta_{l,l'}(t_1,t_2)$, we make use of Eq.~\eqref{eq_evol} and we assume, for sake of comparison with $S_{SQSP}$ in Eq.~\eqref{eq_S_sqsp}, that all system-probe couplings are chosen to be the same. We substitute $t_3$ in Eq.~\eqref{eq_evol} by the upper bound of Eq.~\eqref{eq_condition_t3_main_text}, meaning that we choose $t_3$ in a PQP simulation such that it saturates this upper bound. Since $\text{Var}[\hat{Y}_{pr}] = O(\varepsilon^{2})$, using the above procedure we get $\text{Var}[\Re\{\beta_{l,l'}(t_1,t_2)\}] \sim  O(\varepsilon^2 n_{qub}^2 \mu_{l,l'}^{-2})$, where $\mu_{l,l'}=\sum_{m}\bra{E_{l'}}(\hat{X}_{m}+\hat{Y}_{m})/2\ket{E_{l}}$ is the transition dipole moment of the probed system's energy transition $\ket{E_{l}} \to \ket{E_{l'}}$. We have that $\mu_{l,l'}\leq O(n_{qub})$, hence $\text{Var}[\Re\{\beta_{l,l'}(t_1,t_2)\}] \sim O(\varepsilon^2)$ (and the same for the imaginary part of $\beta_{l,l'}(t_1,t_2)$). Therefore, to estimate $\Re\{\beta_{l,l'}(t_1,t_2)\}$ (or the imaginary part) in Eq.~\eqref{eq_evol} given a desirable (single-qubit) observable measurement accuracy of $\varepsilon$, the number of required shots scales as follows,
\begin{equation}
    S_{PQP} \sim O(\varepsilon^{-2}). \label{eq_S_pqp}
\end{equation} 
This result should be put in contrast to $S_{SQSP}$ in Eq.~\eqref{eq_S_sqsp}, where the latter scales linearly with system's size.

\begin{figure*}
\centering
\includegraphics[width=0.8\textwidth]{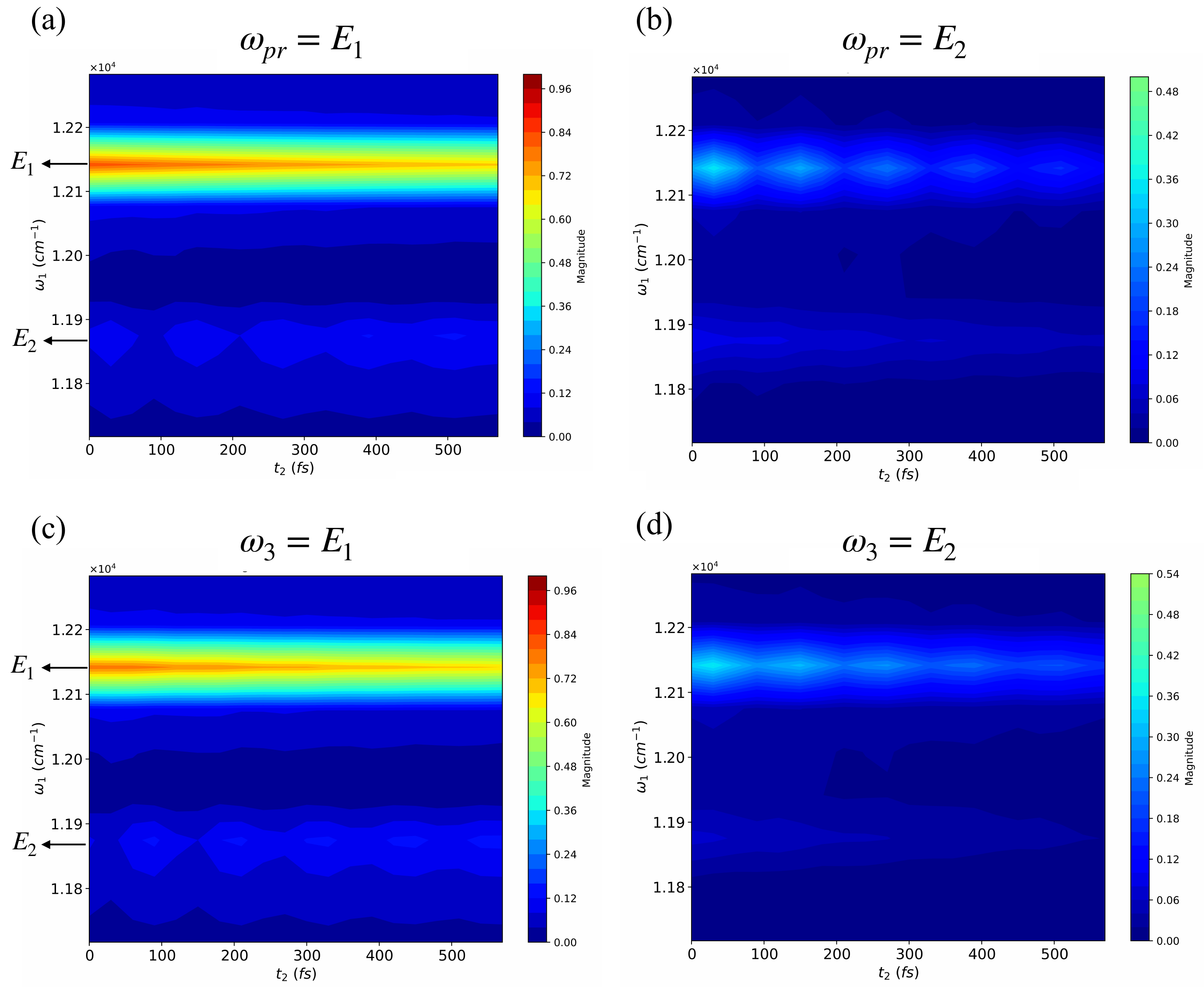}
    \caption{Absolute 2D spectra obtained from rephasing 2DES experiment simulations using the proposed PQP in this work (a,b) and the SQSP (c,d). The system energy eigenvalues are $E_1 \approx 12141$~$cm^{-1}$ and $E_2 \approx 11859$~$cm^{-1}$. We normalize the magnitudes to the interval $[0,1]$ independently for the PQP (a,b) and SQSP (c,d). Here we chose $t_1 = 500$~$fs$ ($\Delta \omega_1 \approx 67$~$cm^{-1}$) and $t_2 = 600$~$fs$ for both PQP and SQSP simulations. We chose different $t_3$ for each protocol, namely, $t_3^{(SQSP)} = t_1$ for the SQSP simulation and $t_3^{(PQP)} = 1.45 t_1$ for the PQP simulation to achieve similar detection frequency resolutions ($\Delta \omega_3 \approx \Delta \omega_1$).}
    \label{fig:results_main_text}
\end{figure*}

With PQP, we can achieve a constant overhead on the number of measurements over the SQSP for a given detection frequency resolution. In comparison to the number of measurements required for the SQSP in Eq.~\eqref{eq_meas_standard}, as long as $N_{freq}<<N_3$, substantial speedups can be achieved with the PQP relatively to the SQSP. The chief cause for this improvement is that the frequency resolution $\Delta \omega_3$ in the PQP is not given by the number of samples, i.e. measurements, as in the SQSP, but by the quantum resource of coherence in the probe qubit. To reconstruct the full 2D spectra, one may choose $N_{freq}=N_3$, recovering Eq.~\eqref{eq_meas_standard} in the worst-case scenario.

Lastly, we show that this novel approach leads to the Heisenberg scaling when applied to peak amplitude estimation on a detection frequency line in a 2D spectra, in contrast to the SQSP which achieves the SQL limit at determining the full 2D spectra. The number of required queries to the Hamiltonian $\hat{H}_{SP}$ implemented during the time interval $t_3$ is given as $Q_{SP}=O(D'_{3})$. Since $D'_{3}$ scales up to a constant factor as $D_3$ in the SQSP (see Sec.~\ref{sec: res_estimation_sqsp}), the estimation of the peak amplitudes in a specific detection frequency line of the 2D spectra with precision $\Delta \omega_3$ is given as,
\begin{equation}
    \Delta \omega_{3} = O\left(\frac{1}{Q_{SP}}\right).
\end{equation}

In Table~\ref{tab:comparison}, we summarize the resource estimate differences between the PQP and SQSP.

\section{Numerical study} \label{sec: numerical}
In this section, we analyze the accuracy of the PQP relatively to the SQSP and its implementation feasibility on quantum hardware. We classically simulate quantum circuits with the SQSP and PQP protocols discussed in Sec.~\ref{sec: standard} and Sec.~\ref{sec: pqp}, respectively. We consider the quantum simulations of a small system of two qubits where instant pulses (i.e. approximate Dirac-delta pulse profiles) are applied to. In the simulations, a second-order Trotter-Suzuki product formula was used with sufficiently small Trotter error and a single-qubit phase-flip probability error of $p_{Z} \approx 1\times 10^{-3}$ was introduced in the system qubits on all Trotter layers. A noiseless probe was added in the PQP simulation. We simulated a two-qubit system with the system Hamiltonian shown in Eq.~\eqref{eq: HS} and we chose Hamiltonian and noise parameters motivated by estimated parameters in photosynthetic systems \cite{caycedo2022exact}, leading to system's energy eigenvalues $E_{1}\approx 12141$~$cm^{-1}$ and $E_{2}\approx 11859$~$cm^{-1}$ in the single-excitation manifold. For more details about the simulations, we refer the reader to Appendix~\ref{sec: simulation_details}.

\subsection{SQSP and PQP simulations}

\begin{figure*}
\centering
\includegraphics[width=0.8\textwidth]{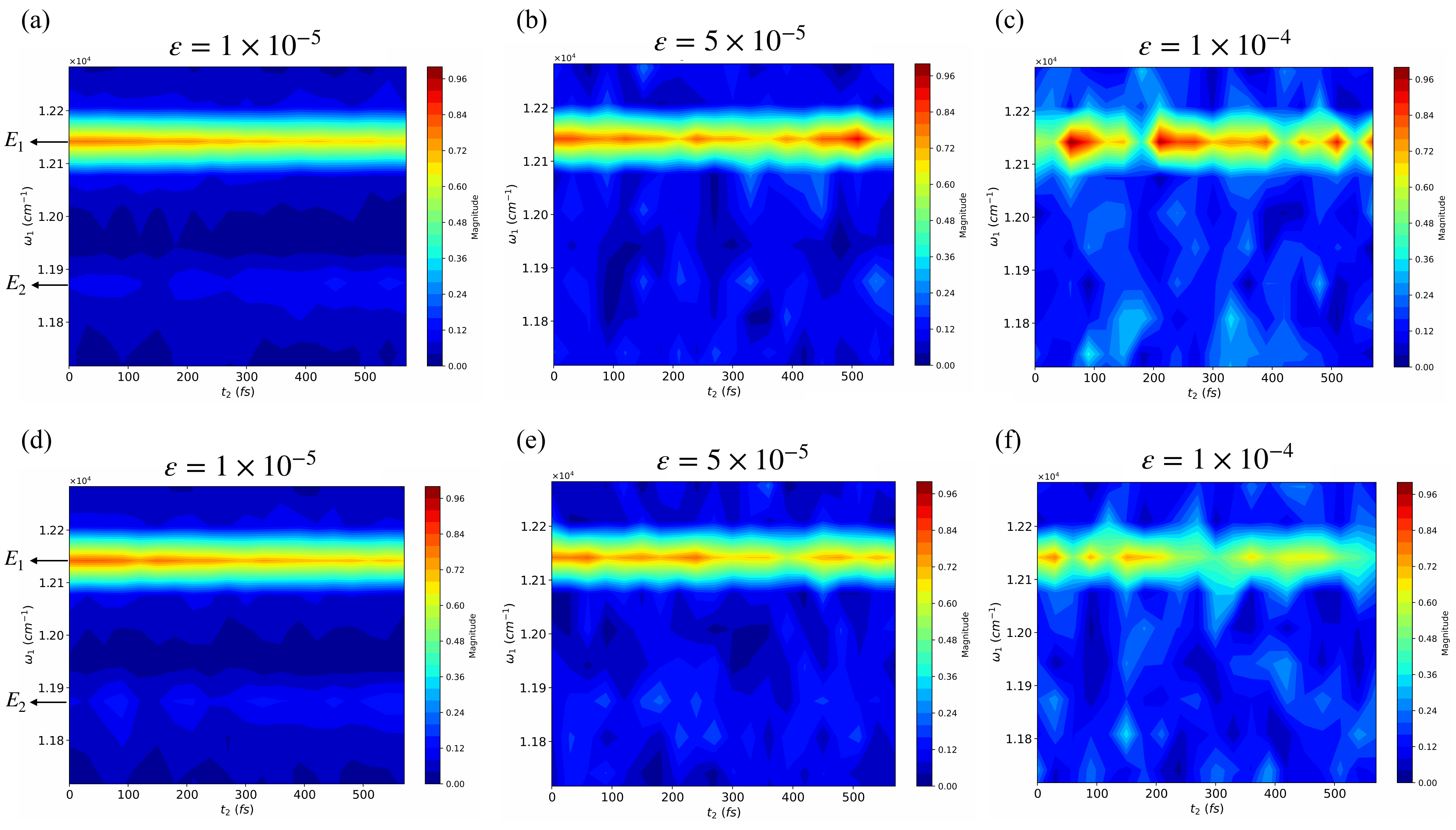}
    \caption{Absolute 2D spectra obtained from rephasing 2DES experiment simulations with shot noise using the proposed PQP in this work (a,b,c) and the SQSP (c,d,e), for $\omega_3 \approx \omega_{pr} = E_1$. For each obtained expectation value in SQSP and PQP simulations shown in Fig.~\ref{fig:results_main_text}, we added a random number sampled from a Gaussian probability distribution with mean $0$ and standard deviation $\varepsilon = 1\times 10^{-5}, 5 \times 10^{-5}$ and $1 \times 10^{-4}$ to mimic the finite number of shots at estimating the expectation values in a realistic quantum simulation scenario. The simulation parameters are the same as in Figure.~\ref{fig:results_main_text}.}
    \label{fig:E1_noisy}
\end{figure*}

\begin{figure*}
\centering
\includegraphics[width=0.8\textwidth]{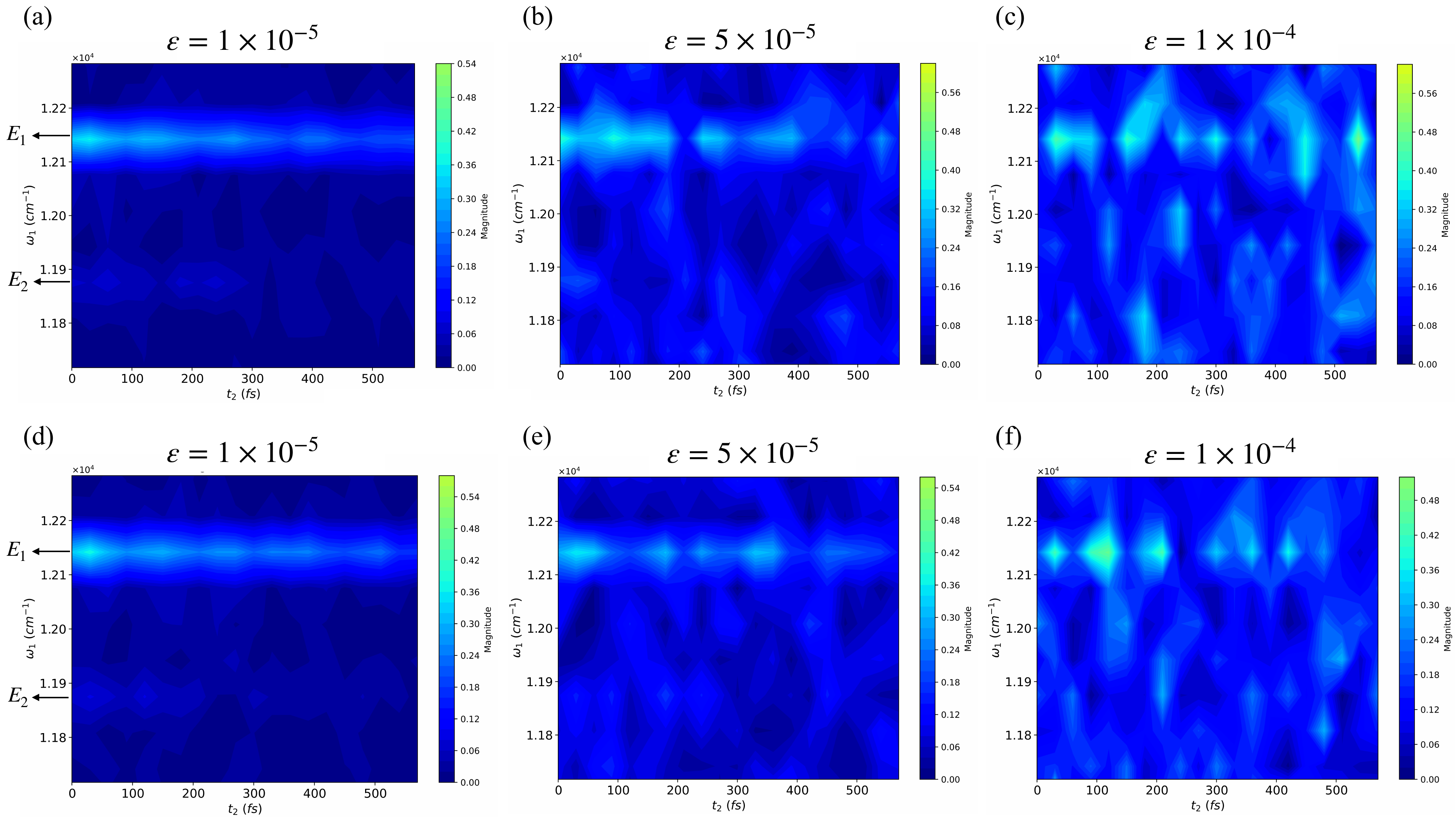}
    \caption{Same as Fig.~\ref{fig:E1_noisy}, but for $\omega_3 \approx \omega_{pr} = E_2$.}
    \label{fig:E2_noisy}
\end{figure*}

In Fig.~\ref{fig:results_main_text}, we show the PQP simulation results for energy gaps of the probe resonant to the ones of the energy eigenvalues of the system, namely, $\omega_{pr}=E_1$ in (a) and $\omega_{pr}=E_2$ in (b). We also show the simulation results obtained with the SQSP in Figs.~\ref{fig:results_main_text}(c) and (d) with the detection frequencies $\omega_3 = E_1$ and $\omega_3 = E_2$, respectively. We observe similar quantitative results between Figs.~\ref{fig:results_main_text}(a) and (c), and Figs.~\ref{fig:results_main_text}(b) and (d), indicating that the probe qubit with energy gap $\omega_{pr}$ effectively behaves as a dynamic filter to a detection frequency $\omega_3$. We also observed that a PQP simulation for a probe energy gap $\omega_{pr} = E_{nres} = E_1 - 500$~$cm^{-1}$, non-resonant with any system's energy gap outputs similar quantitative results as a SQSP simulation (not shown), namely, both simulated 2D spectra contain a constant absolute magnitude $\approx 0$ at $\omega_{pr}\approx \omega_3=E_{nres}$. The reason for this is that no relevant system's dynamics occur in the considered non-resonant energy gap transition, further suggesting that the probe qubit indeed acts as a dynamic detection-frequency filter where $\omega_{pr} \approx \omega_{3}$.

In order to simulate the PQP faithfully, we have chosen the time $t_3^{(PQP)}$ to be larger than that of SQSP, where in the latter $t_3^{(SQSP)}=t_1$ while $t_1$ is the same for both protocol simulations. In particular, in the PQP simulation we chose $t_3^{(PQP)} = 1.45 t_3^{(SQSP)}$. Higher frequency resolutions on the SQSP were not simulated due to its long simulation time. The reason for the different choices of $t_3$ on the PQP and SQSP simulations is that there is an hidden constant factor in the lower bound of the condition of applicability of the PQP presented in Eq.~\eqref{eq_condition_t3_main_text}, hence to achieve the same approximate frequency resolution $\omega_3$ in both simulations, we found that $t_3^{(PQP)}$ must be increased relatively to $t_3^{(SQSP)}$. Furthermore, we confirm that our choices of $t_3$ and system-probe coupling in the PQP simulation (written in the site basis as $J^{(pr)}=J_{1}^{(pr)}=J_{2}^{(pr)} = J_{1,2}/10 = 10$~$cm^{-1}$) are in the weak system-probe interaction regime, which according to Eq.~\eqref{eq_condition_t3_main_text} we find $118<t_3^{(PQP)}< 1667$~$fs$ with our chosen parameters, where $t_3^{(PQP)} = 725$~$fs$. This is also demonstrated by the reasonably good quantitative matching between the SQSP and PQP simulation results in Fig.~\ref{fig:results_main_text}.

We observe small differences between the PQP and SQSP results in Figs.~\ref{fig:results_main_text}(a,c) where $\omega_3 \approx \omega_{pr} = E_1$. In particular, the peak dynamics over $t_2$ in the frequency region $\omega_1 \approx E_2$ for the SQSP and PQP simulations exhibit oscillatory behaviour, which is not clearly shown for the PQP because it has smaller decaying oscillation mean and amplitude than the SQSP. We refer the reader to Appendix~\ref{sec:peak_evolution} for a more informed view on discrepancies of the peak dynamics, where we explicitly plot several peak dynamics of the 2D spectra obtained with the PQP and SQSP simulations. In Figs.~\ref{fig:results_main_text}(b,d), where $\omega_3 \approx \omega_{pr} = E_2$, we also observe a small difference on the maximum magnitude achieved by both SQSP and PQP simulations, which is higher on the latter by a few percentage points. We also observe that the dynamics along $t_2$ of $\omega_1 = E_2$ have overall higher magnitude in the PQP than in the SQSP by about $30$~\%. Overall, we observe small quantitative discrepancies between the peak dynamics of the PQP and SQSP results. As shown in Appendix~\ref{sec:peak_evolution}, we find that slightly changing $J^{(pr)}t_3$ leads to changes on the peak dynamics of the 2D spectra. This suggests that features of the 2D spectra with smaller magnitude are more sensitive to the choice of $J^{(pr)}t_3$ in the PQP simulation, as observed in Fig.~\ref{fig:results_main_text} for $\omega_3 = E_2$. We estimate in Appendix~\ref{sec:peak_evolution} that reasonably accurate results can be obtained in the interval $J^{(pr)}t_3 \in [0.15, 0.22]$ and that the value $J^{(pr)}t_3 \sim 0.22$ provides the smallest overall error on all peak dynamics as shown in Fig.~\ref{fig:results_main_text}. We note that fine-tuning $J^{(pr)}t_3$ within the above interval may lead to more accurate results of the peak dynamics of PQP relatively to the SQSP.

\subsection{Shot noise} \label{sec:shot_noise}

In a realistic quantum simulation scenario, the expectation value estimation is limited by shot noise. In what follows, we quantify the maximum amount of shot noise allowed on each observable measurement to obtain reliable information about the dynamics of the 2D spectra for the PQP and SQSP simulations shown in Fig.~\ref{fig:results_main_text}. More specifically, we add a random number sampled over a Gaussian probability distribution with zero mean and standard deviation $\varepsilon$ to each measured expectation value. We then perform the usual post-processing as explained Sec.~\ref{sec: standard} (Sec.~\ref{sec: pqp}) for the SQSP (PQP) simulation.

In Fig.~\ref{fig:E1_noisy} and Fig.~\ref{fig:E2_noisy} we plot the obtained 2D noisy spectra for $\omega_3 \approx \omega_{pr} = E_1$ and $\omega_3 \approx \omega_{pr} = E_2$, respectively, for different standard deviations $\varepsilon = 1\times 10^{-5}, 5\times 10^{-5}$ and $1 \times 10^{-4}$. We observe that the SQSP and PQP behave similarly, meaning that the accuracy of estimating peak amplitudes with both protocols in the finite-shot regime is equivalent, but limited for the two-qubit system simulations in this work. In particular, variations in the expectation value estimation of $\varepsilon = 1\times 10^5$ on the SQSP and PQP slightly distort the information about the dynamics of the peaks $(\omega_1 = E_2, \omega_3 = E_1)$ as shown in Figs.~\ref{fig:E1_noisy}(a,d) and $(\omega_1 = E_2, \omega_3 = E_2)$ as shown in Figs.~\ref{fig:E2_noisy}(a,d). On the other hand, salient features of the 2D spectra, such as the dynamics of peak amplitudes at $(\omega_1 = E_1, \omega_3 = E_2)$ and $(\omega_1 = E_1, \omega_3 = E_1)$ (see Fig.~\ref{fig:results_main_text} for a comparison) can still be accurately observed with $\varepsilon = 1\times 10^{-5}$. Increasing the shot noise, namely to $\varepsilon = 5\times 10^{-5}$, make the true dynamics of the peaks $(\omega_1 = E_2, \omega_3 = E_1)$ in Figs.~\ref{fig:E1_noisy}(a,d) and $(\omega_1 = E_2, \omega_3 = E_2)$ in Figs.~\ref{fig:E2_noisy}(a,d) to be unobservable. Moreover, the increase of shot noise also introduces irregular oscillatory behaviour on the true dynamics of the cross-peak $(\omega_1 = E_1, \omega_3 = E_2)$ along $t_2$ as shown in Figs.~\ref{fig:E2_noisy}(b,e), as well as in the peak $(\omega_1 = E_1, \omega_3 = E_1)$ as shown in Figs.~\ref{fig:E1_noisy}(b,e). Increasing the shot noise even further to $\varepsilon = 1\times 10^4$, severely distorts the information about the true dynamics on all the considered peaks of the 2D spectra as shown in Figs.~\ref{fig:E1_noisy}(c,f) and Figs.~\ref{fig:E2_noisy}(c,f).

The above results suggest that salient features of the 2D spectra, such as the dynamics of the peaks $(\omega_1 = E_1, \omega_3 = E_2)$ and $(\omega_1 = E_1, \omega_3 = E_1)$, may be observed with variations of $\varepsilon \leq 1\times 10^{-5}$ in the estimation of the expectation values, leading to a required number of shots of $S \geq 10^{10}$ for both SQSP and PQP. We remark that the number of shots scales differently for both protocols as the system's size increases. In particular, a constant (linear) scaling of the number of shots with system's size is expected, as estimated in Eq.~\eqref{eq_S_pqp} (Eq.~\eqref{eq_S_sqsp}), for the PQP (SQSP), thus we expect that the implementation of PQP to simulate 2DES on a quantum device will be more favorable than the SQSP for larger system sizes. Recent progress on improving quantum computer's clock speed as reported by IBM \cite{IBMblog} has shown that superconducting quantum devices can reach nowadays $1.5\times 10^{5}$ circuit layer operations per second (CLOPS), which suggests that $\approx 1.296\times 10^{10}$ circuit layer operations can be achieved in one day. Since $\hat{T}_{k}^{(S)}$ in Eq.~\eqref{eq_Us_trotter} and $\hat{T}_{k}^{(I)}$ in Eq.~\eqref{eq_Ui_trotter} can be implemented in a brick-wall structure \cite{guimaraes2024digital}, they have a constant number of circuit layers as a function of system's size, whereas the implementation of $\hat{T}_{k}^{(SP)}$ in Eq.~\eqref{eq_Usp_trotter} requires a number of circuit layers scaling linearly with system's size due to the qubit-probe interaction. Therefore, we estimate that on our simulations one would need a \emph{maximum} of $C_{max}^{(PQP)} \sim 10^2 n_{qub}$ ($C_{max}^{(SQSP)}\sim 10^3$) circuit layers on a single quantum circuit and $\sim 10^{10}$ ($\sim 10^{10}n_{qub}$) shots to estimate the observable in Eq.~\eqref{eq_evol} (Eq.~\eqref{eq_F_obs}) in the quantum circuit to a good accuracy using the PQP (SQSP) for a given $(t_1,t_2,\vec{\phi}_{k})$, leading to a total maximum of $\sim 10^{12}n_{qub}$ ($\sim 10^{13}n_{qub}$) circuit layers. We envision that further advances on quantum hardware, such as the increase of the clock speed (i.e. CLOPS), and quantum algorithm development, e.g. in view of achieving a reduction of the effect of shot noise in the accuracy of the PQP simulation, can lead to a reduction of several orders of magnitude on the resource requirements to implement 2DES simulations on quantum computers using PQP (see Sec.~\ref{sec:conclusion} for future work). Furthermore, we highlight here the potential use of randomized product formulas \cite{nakaji2023qswift,nakaji2024high}, which can eliminate the system's size dependence on the number of circuit layers in the PQP simulation at the cost of a higher sampling overhead.

\subsection{Classical simulations}

We observe that the classical simulations performed in Fig.~\ref{fig:results_main_text} with PQP are substantially faster and require less memory to store the expectation values than the SQSP. By classical simulations, we mean simulations of the quantum circuits in a classical computer followed by expectation value estimation in the infinite-shot regime. We observed that in our simulations, executed in a 30-core, 2.1GHz base frequency Intel Xeon E6252 Gold, the PQP requires $9.5$~$h$ to obtain the required expectation values to estimate peak amplitudes on one detection frequency line, whereas the SQSP requires $14$~$h$ to estimate the full 2D spectra. Therefore, the PQP requires $\approx 68$~\% of the SQSP execution time. Regarding the memory requirements for each simulation, we observed that storing the SQSP expectation values takes a total of $\approx 2$~$GB$ for the two-qubit example of Fig.~\ref{fig:results_main_text}, whereas the PQP takes $\approx 3.5$~$MB$ to store the expectation values to estimate the absolute peak amplitudes of one detection frequency line, leading to three orders of magnitude of memory savings for the PQP relatively to SQSP. We expect that 2DES simulations with longer $t_2$ and with higher frequency resolutions $\omega_1$ and $\omega_3$ can bring even larger memory savings and reduced simulation execution times. On the other hand, for larger system sizes, classical simulations typically face trouble due to the long timescales involved in 2DES experiments \cite{engel2007evidence,thyrhaug2018identification}, hence in such regime quantum simulations may provide an advantage on simulation runtime and memory.

\subsection{Resource estimate for a 2DES simulation applied to the FMO complex}
We now estimate the computational resources required in a digital quantum simulation of 2DES using the SQSP and PQP to replicate a recent 2DES experiment \cite{thyrhaug2018identification} performed in the Fenna-Matthews-Olson (FMO) complex in green sulphur bacteria. The FMO complex can be simulated with eight light-harvesting molecules structured in an approximate 2D network \cite{tronrud2009structural}. We considered the parameters used in the experiment in Ref.~\cite{thyrhaug2018identification} with Gaussian pulses up to $14$~$fs$ (see Appendix~\ref{sec:resource_details}). We assume that a quantum hardware platform with all-to-all qubit connectivity such as neutral atom or trapped-ion quantum devices is available to us and we considered the implementation of phase-cycling to resolve the real or imaginary part of a rephasing 2D spectra signal \cite{thyrhaug2018identification}. We calculated the computational resources for two detection frequency lines in the PQP simulation, similar to the ones analyzed in Ref.~\cite{thyrhaug2018identification}. We refer the reader for Appendix~\ref{sec:resource_details} for more details.

\begin{figure}
\centering
\includegraphics[width=0.47\textwidth]{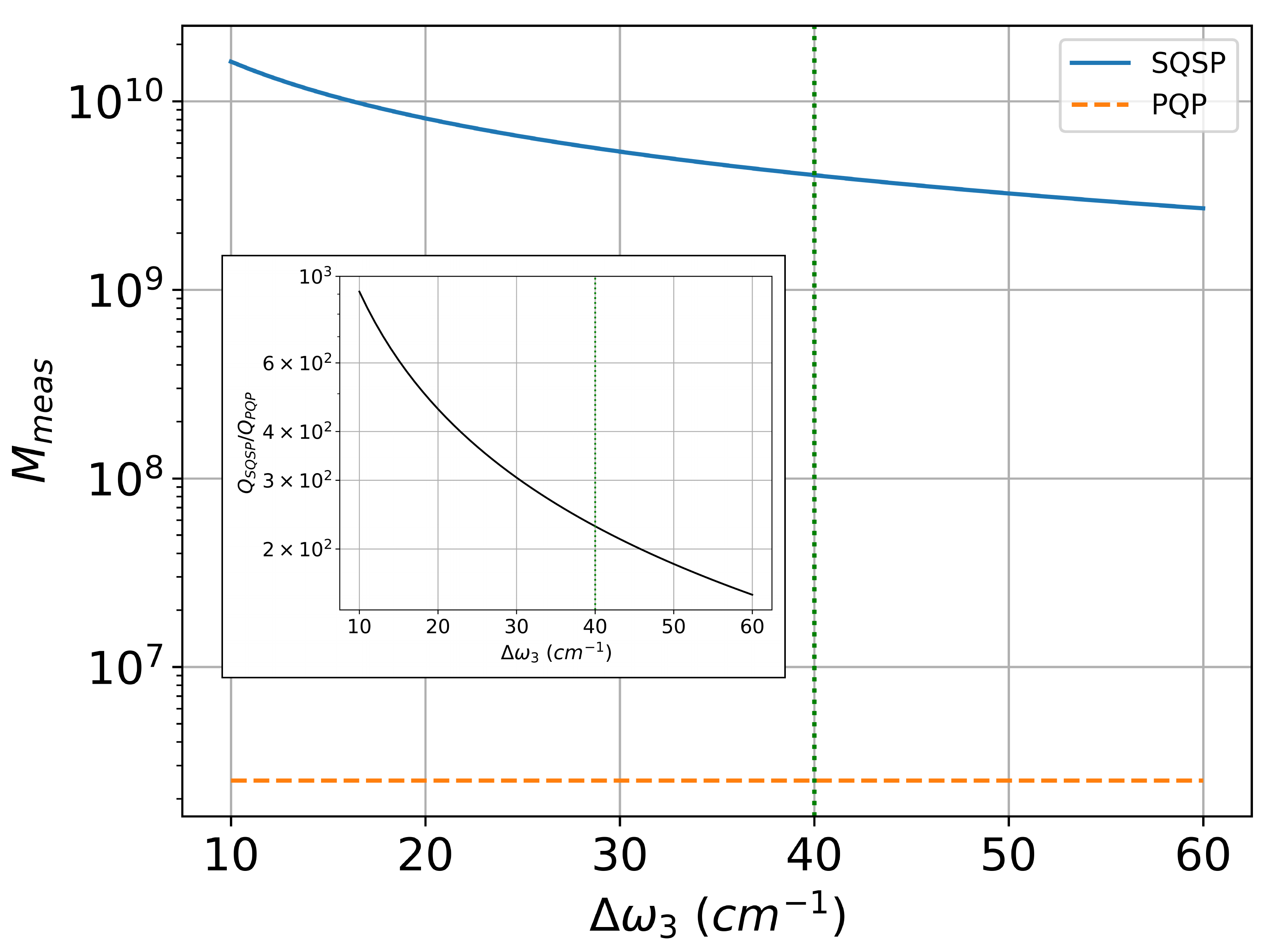}
    \caption{Number of measurements $M_{meas}$ required for the SQSP and PQP simulations of a 2DES experiment of the FMO complex \cite{thyrhaug2018identification} as a function of the detection frequency resolution $\Delta \omega_3$ in the infinite-shot regime. In the inset, we show the ratio between simulation runtimes $Q_{SQSP}$ ($Q_{PQP}$) of the SQSP (PQP) or, equivalently, the ratio between the number of queries to the system Hamiltonian of both protocols. The vertical line denotes the detection frequency resolution $\Delta \omega_3 = 40$~$cm^{-1}$ considered in the experiment of Ref.~\cite{thyrhaug2018identification}.}
    \label{fig: fmo}
\end{figure}

In Fig.~\ref{fig: fmo}, we illustrate the number of measurements of the PQP and SQSP simulations (in the infinite-shot regime). As it can be observed, a reduction of about three orders of magnitude in the number of measurements of the PQP relatively to the SQSP can be obtained for a detection frequency resolution in the range $\Delta \omega_3 \in [10,60]$~$cm^{-1}$, which decreases as one increases the detection frequency resolution $\Delta \omega_3$. The PQP reduction in the number of measurements relatively to the SQSP is particularly large for high-accuracy simulations, i.e. for small $\Delta \omega_3$. We also show the estimated simulation runtime speedup for the PQP and SQSP simulations, i.e. the ratio of system Hamiltonian queries $Q_{SQSP}/Q_{PQP}$, in the inset of Fig.~\ref{fig: fmo}. We observe a reduction of the simulation runtime by more than two orders of magnitude of the PQP relatively to the SQSP at $\Delta \omega_3 = 40$~$cm^{-1}$, as considered in the experiment of Ref.~\cite{thyrhaug2018identification} and identified by the vertical dashed line. Such speedup comes at the expense of the PQP requiring a higher number of Trotter layers than SQSP, namely $D_{max}^{(PQP)} \geq 1403 \approx 1.14 D_{max}^{(SQSP)}$. We remark that accounting for the finite number of shots yields an increase of $10^{10}$ to the number of measurements shown in Fig.~\ref{fig: fmo} on both protocols following the results obtained in Sec.~\ref{sec:shot_noise}.

%Furthermore, in order to resolve the expectation values of Eq.~\eqref{eq_evol} in the PQP, we require their distributions to have reasonably high standard deviations $\sigma$, as discussed in the previous section. Following Eq.~\eqref{eq_p_dep}, we may obtain the maximum depolarization error probability $p_{dep} = \gamma_{dep}t_3/D_3$ on each Trotter layer allowed by the quantum device as a function of $\sigma$. Therefore, using the fitting parameters found in Fig.~\ref{fig: noise} and choosing a desirable standard deviation of $\sigma_{\rm sampl} = \sigma/5 = 10^{-3}$, we find a maximum error probability of $p_{dep} \leq 8.12 \times 10^{-6}$ per Trotter layer in the PQP simulation with a required number of shots per circuit run of $S=10^{6}$ which may be feasible in early fault-tolerant quantum computers~\cite{acharya2024quantum}, where quantum error correction protocols can help to achieve lower error rates \cite{google2023suppressing}.

\section{Conclusion} \label{sec:conclusion}
In this work, we introduced the probe qubit protocol (PQP) for simulating two-dimensional electronic spectroscopy (2DES) experiments on quantum computers. By integrating phase-cycling into the simulation, the PQP is applicable to systems of any size. The protocol involves attaching a probe qubit to the quantum circuit with connections to all system qubits during the detection stage of 2DES. Compared to the SQSP, the PQP offers computational advantages, including Heisenberg-scaling on detection frequency resolution and requiring only a single-qubit measurement per circuit run, regardless of system size. This results in significant speedups, particularly when probing a limited number of detection frequency lines, which is often the case in applications like energy transport studies in photosynthetic systems.

We have outlined the conditions for implementing the PQP on near-term quantum hardware, such as neutral atom or trapped-ion devices, which can already support high-fidelity two-qubit interactions and single Pauli measurements on the probe qubit at the end of the quantum circuit. While shot noise and decoherence are detrimental to the probe qubit protocol, increasing the circuit throughput of the quantum device and improving the PQP algorithm may mitigate the former, whereas employing quantum error correction or mitigation techniques can reduce the effect of the later on the accuracy of the simulation.

The PQP presents a promising application for quantum computers, particularly in simulating 2DES with realistic quantum systems \cite{guimaraes2023noise,guimaraes2024digital} where classical simulations face challenges, such as in noisy energy transport in photosynthetic networks \cite{caycedo2022exact,lorenzoni2024systematic}. Future work could enhance the protocol by adding an extra probe qubit to filter specific excitation frequency $\omega_1$ evolutions, further reducing the cost of 2DES simulation implementations when only a few peak amplitudes in the 2D spectra need to be estimated. Notably, using a matrix completion method \cite{AlmeidaPP12,scheuer2015accelerated}, the full 2D spectra can be reconstructed without probing every detection frequency, offering an even greater speedup over standard protocols. We expect these improvements to reduce the number of circuit layers of the PQP by a few orders of magnitude. The development of better measurement protocols, e.g. with additional probes, are also a promising direction towards reducing the effect of shot noise in the accuracy of the 2DES simulation using PQP, e.g. via achieving Heisenberg scaling on expectation value estimation, i.e. $S_{PQP}= O(1/\varepsilon)$. This would additionally reduce the number of required circuit layers by a few orders of magnitude. Applying this protocol to unsolved problems in real quantum devices, like energy transport in the Fenna-Matthews-Olsen complex \cite{caycedo2022exact}, could demonstrate quantum computers' capabilities in tackling problems beyond classical reach. This could be achieved by using Quantum TEDOPA \cite{guimaraes2024digital} to simulate open system dynamics or analog approaches on trapped-ion quantum devices \cite{lemmer2018trapped,macdonell2021analog}. One other possible future improvement venue would be to tailor the proposed probe qubit protocol to other spectroscopy experiments, such as for instance, Multiple-Quantum Nuclear Magnetic Resonance (MQ-NMR) \cite{feike1996broadband}.

After completion of this work, we became aware of Ref.~\cite{stenger2022simulating} which examines the use of a probe qubit to compute one-dimensional spectra.

\begin{acknowledgments}
JDG acknowledges funding from the 
Portuguese Foundation for Science and Technology (FCT) through PhD grant UI/BD/151173/2021. JDG, JL, MBP and SFH acknowledge support by the BMBF 
project PhoQuant (grant no. 13N16110). MIV acknowledges support from the FCT through Strategic Funding UIDB/04650/2020. SFH and MBP acknowledge
support by the DFG via the QuantERA project ExtraQt. The authors acknowledge support by the state of Baden-Württemberg through bwHPC and the 
German Research Foundation (DFG) through Grant No. INST 40/575-1 FUGG (JUSTUS 2 cluster). JDG and MIV are thankful to M. Belsley for elucidating discussions.
\end{acknowledgments}

\appendix

\section{Evolution operator over time $t_3$}\label{sec: probe_system_dynamics}
Let us rewrite the probe-system interaction Hamiltonian in Eq.~\eqref{eq: HPR} in the system's energy eigenbasis.
We define $\hat{a}=(\hat{X}+i\hat{Y})/2$ ($\hat{a}^{\dag}=(\hat{X}-i\hat{Y})/2$) as an excitonic annihilation (creation) operator acting on the probe qubit or system qubit and rewrite Eq.~\eqref{eq: HPR} as follows,
\begin{equation}
    \hat{H}_{PR} = \omega_{pr}\hat{a}^{\dag}_{pr}\hat{a}_{pr} + \sum_{m} J_m^{(pr)} (\hat{a}^{\dag}_{pr}\hat{a}_{m} +\hat{a}^{\dag}_{m}\hat{a}_{pr}). \label{eq_HPR_exc}
\end{equation}
Now we change to the energy eigenstates, $\ket{E_{k}}$, i.e. the eigenbasis of $\hat{H}_{S}$ given by Eq.~\eqref{eq: HS}. The site operators are transformed as $\hat{a}_{m} = \sum_{k',k} \alpha_{k',k}^{(m)}\ket{E_{k'}}\bra{E_{k}}\equiv \sum_{k',k} \alpha_{k',k}^{(m)}\hat{A}_{k',k}$, where $\alpha_{k',k}^{(m)} = \bra{E_{k'}}\hat{a}_m \ket{E_{k}}$ and $\hat{A}_{k',k}$ changes the energy state from $\ket{E_{k}}$ to $\ket{E_{k'}}$. A similar transformation takes place for $\hat{a}^{\dag}_{m}$. 

Since $\hat{a}_{m}$ destroys an exciton in the system, $\ket{E_{k}}$ and $\ket{E_{k'}}$ span different excitation manifolds. The ground-state, single- and double-excitation manifolds are represented by $\mathcal{M}_{0}$, $\mathcal{M}_{1}$ and $\mathcal{M}_{2}$, respectively. Higher-excitation manifolds can also be considered. Following the above reasoning, we now fix, without loss of generality, $k'<k$ and $k''>k$.

Applying the above change of basis to the system's creation and annihilation operators in Eq.~\eqref{eq_HPR_exc}, we arrive to the following system-probe interaction Hamiltonian,
\begin{align}
    \hat{H}'_{PR} = & \omega_{pr}\hat{a}^{\dag}_{pr}\hat{a}_{pr} \nonumber \\
    & + \sum_{p}\sum_{k\in \mathcal{M}_{p}}
     \hat{a}^{\dag}_{pr}\left(\sum_{k'\in \mathcal{M}_{p-1}} J_{k',k}^{(pr)} \hat{a}_{k',k} \right) \nonumber \\
    &+ \left(\sum_{k''\in \mathcal{M}_{p+1}} J_{k,k''}^{(pr)*} \hat{a}^{\dag}_{k,k''} \right)\hat{a}_{pr}. ~\label{eq_H'pr}
\end{align}
Here, $J_{k',k}^{(pr)} = \sum_{m}J_{m}^{(pr)}\alpha_{k',k}^{(m)}$ denotes the coupling strength between the probe and the system. This equation suggests that when we evolve the system and probe with  $\hat{H}'_{PR}$, jumps between system's energy eigenstates between different adjacent manifolds may occur. 

To further understand the effect of such interaction on the system over a time $t_3$, we split the Hamiltonian in Eq.~\eqref{eq_H'pr} into $\hat{H}''_{PR} = \hat{H}_{0}+ \hat{H}_{int}$, where $\hat{H}_{0} = \hat{H}_S + \omega_{pr}\hat{a}^{\dag}_{pr}\hat{a}_{pr}$ and $\hat{H}_{int} = \hat{H}'_{PR}-\omega_{pr}\hat{a}^{\dag}_{pr}\hat{a}_{pr}$. With this new notation, we move $\hat{H}''_{PR}$ to the interaction picture, such that,
\begin{align}
        &\hat{H}''_{PR}(t) = e^{i\hat{H}_{0}t}\hat{H}_{int}e^{-i\hat{H}_{0}t} \nonumber \\
         &= \sum_{p}\sum_{k\in \mathcal{M}_{p}}
        \hat{a}^{\dag}_{pr}\left(\sum_{k'\in \mathcal{M}_{p-1}} e^{i(\omega_{pr}-\omega_{k,k'})t}J_{k',k}^{(pr)} \hat{a}_{k',k} \right) \nonumber\\
        &+ \left(\sum_{k''\in \mathcal{M}_{p+1}} e^{-i(\omega_{pr}-\omega_{k'',k})t}J_{k,k''}^{(pr)*} \hat{a}^{\dag}_{k,k''} \right)\hat{a}_{pr},\label{eq_time_HPR}
\end{align}
where $\omega_{k,k'} = E_{k}-E_{k}'$ and $E_{k}$ is the energy of the energy eigenstate $\ket{E_{k}}$ of the system. In order to obtain the operator $\hat{U}'_{SP}(t)$ that evolves the system and probe qubits over time $t_3$ with the Hamiltonian $\hat{H}''_{PR}(t)$, we make use of the first-order Magnus expansion~\cite{Magnus}, yielding the evolution operator
\begin{equation}
    \hat{U}'_{SP}(t_3) \approx \exp{-i\int_{0}^{t_3} dT\hat{H}''_{PR}(T)}. \label{eq_integral}
\end{equation}
This approximation is valid if the probe-system couplings 
$J_{k',k}^{(pr)}$ are kept small, such that we can ignore terms like $\int \int dT_1 dT_2[\hat{H}''_{PR}(T_1),\hat{H}''_{PR}(T_2)]/2$ and other higher-order commutators. The next leading term, i.e. the second-order term of the Magnus expansion is as follows $\int \int dT_1 dT_2 [\hat{H}''_{PR}(T_1),\hat{H}''_{PR}(T_2)]$. We now derive its asymptotic scaling as follows. Since the evolution frequencies of each Hamiltonian shown in Eq.~\eqref{eq_time_HPR} contained in the product of this commutator must be of opposite sign in order not to vanish when we apply the integrals in the long time-limits $T_1$ and $T_2$ (i.e. via the rotating-wave approximation), we have that the resulting number of energy transitions with energy gap $\Delta E=\omega_{k,k'}$ allowed in the worst-case, i.e. number of terms, is $O(N_{\Delta E}^2)$. The second-order term in the Magnus expansion then scales as $O(N_{\Delta E}^2\max_{a',a,b',b}[J_{a',a}^{(pr)}J_{b',b}^{(pr)}])$. Since the highest-excitation manifold of the system, which we call $\mathcal{M}_{\max{(p)}}$, has the highest number of possible states among all considered manifolds, namely $n_{qub}^{\max(p)}$ states, and in $\mathcal{M}_{\max{(p)-1}}$ we have $n_{qub}^{\max(p)-1}$ states, a maximum of $N_{\Delta E}< O(n_{qub}^{\max(p)-1})$ energy transitions are possible between the two highest-excitation manifolds, i.e.  $\mathcal{M}_{\max{(p)-1}}$ and $\mathcal{M}_{\max{(p)}}$. Regarding the couplings $J_{k',k}^{(pr)}$, we have that $ J_{k',k}^{(pr)} = \sum_{m}J_{m}^{(pr)}\alpha_{k',k}^{(m)}$, hence the maximum value each $J_{k',k}^{(pr)}$ can take is $\max_{m}[J_{m}^{(pr)}]$. Using the above parameters, the second-order term of the Magnus expansion shown above scales as $\int \int dT_1 dT_2[\hat{H}''_{PR}(T_1),\hat{H}''_{PR}(T_2)] < O\left(n_{qub}^{2(\max(p)-1)}\max_{m}[J_{m}^{(pr)}]^2\right)$ in the long-time limit. Assuming that excitations in the system build up only up to the double-excitation manifold $\max(p)=2$, we may choose sufficiently small system-probe dipolar couplings scaling as $\max_{m}[J_{m}^{(pr)}]\leq O(n_{qub}^{-1})$, in order to neglect the second-order term of the Magnus expansion. We note that considering excitations up to $\max(p)=2$ in the system is typically assumed in simulations of photosynthetic systems due to their properties and structure \cite{mohseni2014quantum}.

If we let $t_3$ be sufficiently long in comparison to the inverse of  $\omega_{pr,s,s'}=\min(|\omega_{pr}-\omega_{s,s'}|)$, where $\omega_{s,s'}=E_{s}-E_{s'}$ is the energy gap of the system closest to $\omega_{pr}$ and $E_{s}$ is the energy of the system's energy eigenstate $\ket{E_{s}}$, we have a relative long-time limit of $t_3>O(2\pi \omega_{pr,s,s'}^{-1})$. In this case, phases with frequencies $\omega_{k,k'} \neq \omega_{pr}$ and $\omega_{k'',k} \neq \omega_{pr}$ in Eq.~\eqref{eq_integral} with Eq.~\eqref{eq_time_HPR} approximately average to $0$, hence the integral in Eq.~\eqref{eq_integral} can be approximated as,
\begin{align}
    \hat{U}'_{SP}(t_3)  \approx  & \exp \left \{-it_3
\sum_{\omega_{k,k'}=\omega_{k'',k}=\omega_{pr}}  \left [ J_{k',k}^{(pr)} \hat{a}^{\dag}_{pr}\hat{a}_{k',k} \right. \right. \nonumber \\
           & \left. \left. +  J_{k,k''}^{(pr)*} \hat{a}^{\dag}_{k,k''}\hat{a}_{pr} \right ]\right\} . \label{eq_approximation_time}
\end{align}

We reorganize the indices so that the approximated evolution operator is expressed as follows,
\begin{align}
    \hat{U}'_{SP}(t_3) \approx &
    \exp\left\{-it_3\sum_{\omega_{l,l'}=\omega_{pr}} \left [J_{l',l}^{(pr)} \hat{a}^{\dag}_{pr}\hat{a}_{l',l}  \right.\right. \nonumber \\
    & \hspace*{3.cm}\left. \left. + J_{l',l}^{(pr)*} \hat{a}^{\dag}_{l',l}\hat{a}_{pr}\right ]\right\},\label{eq_semilast_U_t3}
\end{align}
where we now fixed $l'<l$. Since $\hat{H}_{S}$ and $\hat{a}_{m}$ are real, $J_{l',l}^{(pr)}=J_{l',l}^{(pr)*}$, and we have,

\begin{align}
    \hat{U}'_{SP}(t_3) \approx & \exp\left\{-it_3\sum_{\omega_{l,l'}=\omega_{pr}} J_{l',l}^{(pr)}\left(\hat{a}^{\dag}_{pr}\hat{a}_{l',l}  + \hat{a}^{\dag}_{l',l}\hat{a}_{pr}\right)\right\} .\label{eq_last_U_t3}
\end{align}

Let us expand the approximated evolution operator in Eq.~\eqref{eq_last_U_t3} in a Taylor series up to first-order,
\begin{equation}
    \hat{U}^{(1)}_{SP}(t_3) \approx \hat{I}-it_3 \left(\sum_{\omega_{l,l'}=\omega_{pr}} J_{l',l}^{(pr)} \left[\hat{a}^{\dag}_{pr}\hat{a}_{l',l} + \hat{a}^{\dag}_{l',l}\hat{a}_{pr}\right]\right). \label{eq_U_approx_Taylor}
\end{equation}
We shall now calculate how $t_3$ must scale as a function of system's size and system-probe coupling in order to evaluate the accuracy of the above expansion, while keeping $t_3$ in the long-time limit. In order to employ the first-order Taylor expansion of the evolution operator as we did in Eq.~\eqref{eq_U_approx_Taylor}, we must have 
\begin{displaymath}
    t_3<\left[\sum_{\omega_{l,l'}=\omega_{pr}} J_{l',l}^{(pr)}\right]^{-1} \!\!\!= O\left(\left[N_{\Delta E=\omega_{pr}} \max_{m}[J_{m}^{(pr)}]\right]^{-1}\right).
\end{displaymath}
From the derivation above, we have that $N_{\Delta E} < O(n_{qub})$ if we consider $\max(p)=2$ and long-time limit $t_3$. Therefore, we must choose 
\begin{displaymath}
    t_3< O\left(\left[n_{qub}\max_{m}[J_{m}^{(pr)}]\right]^{-1}\right)    
\end{displaymath} 
to have an accurate first-order expansion of the evolution operator in Eq.~\eqref{eq_U_approx_Taylor}.

This inequality is understandable intuitively. The probe qubit interacts with all qubits representing the molecules in the system, thus the effective interaction parameter between it and the system is $n_{qub}$ times stronger. At the same time, $t_3$ must be long enough to guarantee that the last excitation will propagate among all molecules, i.e. $\min \{J_{mn}t_3\} > 1$, leading to the lower bound in Eq.~(\ref{eq_condition_t3_main_text}).

\section{Measurement of probe coherence terms}\label{sec: coherence_term}
Here, we derive the outcome of measuring the observables $\hat{O}_{pr} = \hat{X}_{pr}$ or $\hat{O}_{pr} = \hat{Y}_{pr}$ in the noiseless probe regime.

The initial density matrix of the system and probe at $t_3=0$ is given as,
\begin{equation}
    \hat{\rho}_{SP}(t_1,t_2) = \sum_{k',k} \beta_{k',k}(t_1,t_2)\ket{E_{k'}}\bra{E_{k}}_{S}\otimes \ket{0}\bra{0}_{pr}.
\end{equation}
Bringing $\hat{\rho}_{SP}$ to the interaction picture yields,
\begin{align}
    \hat{\rho}_{SP}&(t_1,t_2,t_3)   \nonumber\\
    & = \sum_{k',k} e^{i\omega_{k',k}t_3}\beta_{k',k}(t_1,t_2)\ket{E_{k'}}\bra{E_{k}}_S \otimes \ket{0}\bra{0}_{pr}. \label{eq_rho_SP}
\end{align}
We now apply the perturbative evolution operator in Eq.~\eqref{eq_U_approx_Taylor} to the previous density matrix in Eq.~\eqref{eq_rho_SP}. We want to compute the expectation value of $\hat{X}_{pr}$ ($\hat{Y}_{pr}$) in the perturbative system-probe interaction regime. Therefore, we calculate the effect of the perturbative interaction that leads to non-vanishing coherence terms, $\ket{1}\bra{0}_{pr}$ and $\ket{0}\bra{1}_{pr}$, in the probe's density matrix, which enables us to discard second-order terms in the system-probe couplings. Applying then Eq.~\eqref{eq_U_approx_Taylor} to the density matrix in Eq.~\eqref{eq_rho_SP}, leads to the following result accurate up to the second-order Taylor expansion in Eq.~\eqref{eq_U_approx_Taylor},
\begin{align}
     & \hat{U}^{(1)\dag}_{SP}(t_3) \hat{\rho}_{SP}(t_1,t_2,t_3) \hat{U}^{(1)}_{SP}(t_3) \xrightarrow[\text{coherence 
terms}]{} it_3 \nonumber\\
 & \times \sum_{k,\omega_{l,l'}=\omega_{pr}}e^{it_3\omega_{l,k}}
\beta_{l,k}(t_1,t_2)J_{l',l}^{(pr)}\ket{E_{l'}}\bra{E_{k}}_{S}\otimes \ket{1}\bra{0}_{pr} \nonumber \\
 & + H.c.,
\end{align}

where $l>l'$ and $l=k'$. Performing then the measurement of the observable $\hat{X}_{pr}(t_{3})$ in the interaction picture, one has $l'=k$, yielding the expectation value of $\hat{X}_{pr}(t_{3})$ as,
\begin{align}
     & \Tr[\hat{X}_{pr}(t_{3}) \hat{U}^{(1)\dag}_{SP}(t_{3}) \hat{\rho}_{SP}(t_1,t_2,t_3)\hat{U}^{(1)}_{SP}(t_3)] \nonumber \\
    & \hspace*{2cm}= 2t_3 \sum_{\omega_{l,l'}=\omega_{pr}} J_{l',l}^{(pr)}\Im[\beta_{l',l}(t_1,t_2)].\label{eq_appendix_X_exp}
\end{align}
By performing the procedure outlined here, the perturbative system-probe interaction extracts information from the imaginary component of coherence terms $\beta_{l',l}(t_1,t_2)\ket{E_{l'}}\bra{E_{l}}$, where the energy gap between states $\ket{E_{l}}$ and $\ket{E_{l'}}$ is $\omega_{l,l'}=\omega_{pr}$.

Performing the measurement of $\hat{Y}_{pr}(t_{3})$ instead results in the expectation value,
\begin{align}
& \Tr[\hat{Y}_{pr}(t_{3})\hat{U}^{(1)\dag}_{SP}(t_{3})\hat{\rho}_{SP}(t_1,t_2,t_3)\hat{U}^{(1)}_{SP}(t_3)] \nonumber \\
&\hspace*{2cm} = 2t_3\sum_{\omega_{l,l'}=\omega_{pr}} J_{l',l}^{(pr)}\Re[\beta_{l',l}(t_1,t_2)].\label{eq_appendix_Y_exp}
\end{align}

\section{Simulation details}\label{sec: simulation_details}
In this section, we describe the the simulations presented in Sec.~\ref{sec: numerical}.

\begin{figure*}
\centering
\includegraphics[width=0.8\textwidth]{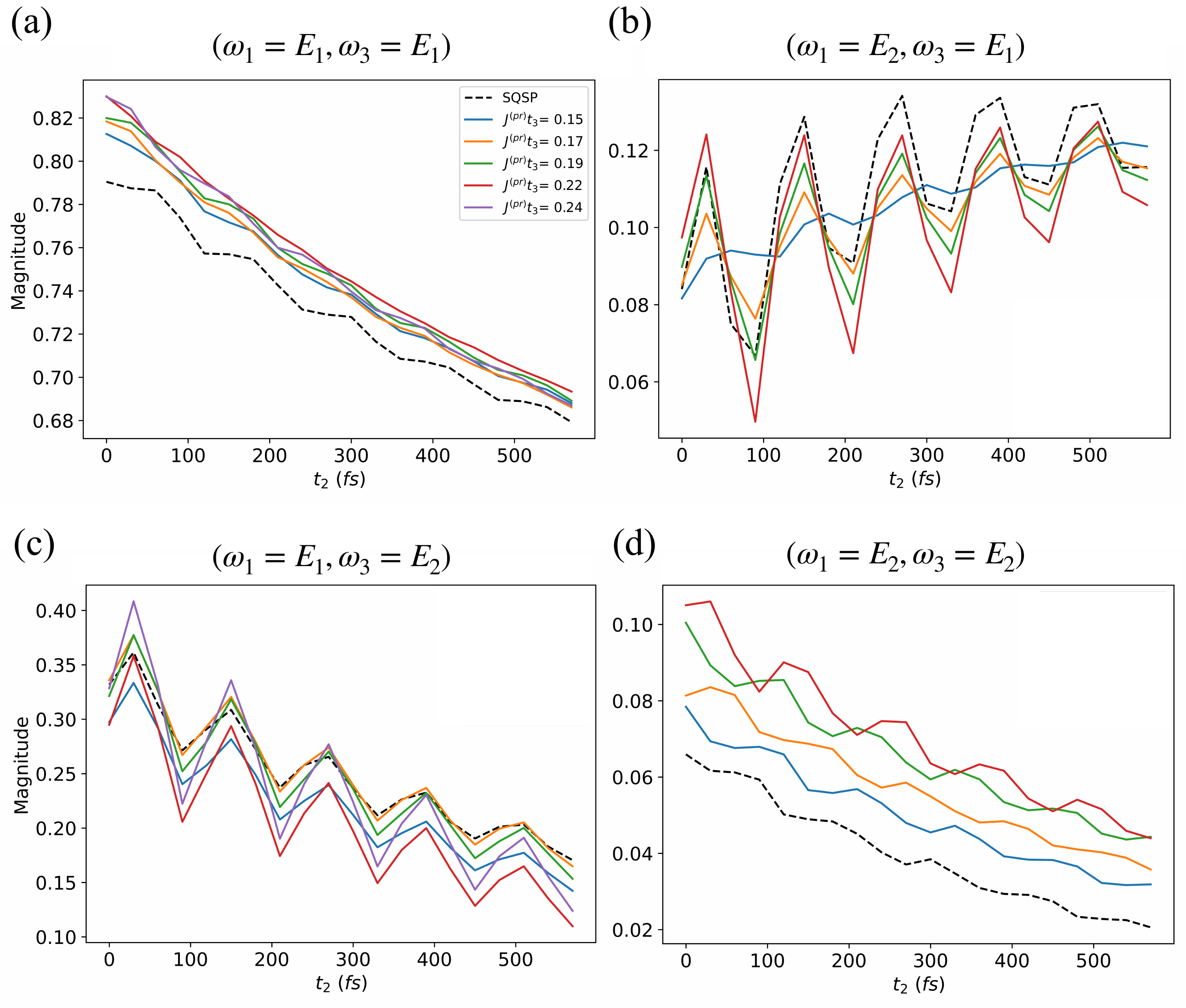}
    \caption{Dynamics of the absolute amplitude of peaks over $t_2$ in the rephasing 2D spectra with SQSP, and the PQP for several choices of $J^{(pr)}t_3 = 0.15$, $0.17$, $0.19$, $0.22$ and $0.24$. We plot the dynamics with 20 sampled data points over $t_2$, thus the lines serve merely as guides to the eye.}
    \label{fig:peak_evol}
\end{figure*}

We considered two molecules (i.e. qubits) as the system. The molecular excited-state energies are $E^{(mol)}_1=12100$~$cm^{-1}$ and $E^{(mol)}_2=11900$~$cm^{-1}$ as in typical photosynthetic systems \cite{mohseni2014quantum}. The intermolecular coupling strength written in the site basis is $J_{1,2} = 100$ $cm^{-1}$. This system has energy eigenvalues $E_{1}\approx 12141$~$cm^{-1}$ and $E_{2}\approx 11859$~$cm^{-1}$ in the single-excitation manifold.

All quantum circuit simulations were developed in Python with the \emph{Cirq} library. We implement the second-order Trotter-Suzuki product formula as the time-evolution method.
 During the interval of time $t_1 = 500$~$fs$, we considered a total of $D_{1} = 400$ Trotter iterations with a number $N_1 = D_1$ of data samples, reaching a frequency resolution of $\Delta\omega_{1}\approx 66.71$~$cm^{-1}$, in the SQSP and PQP simulations. During the interval of time $t_2 = 600$~$fs$, we considered $N_2 = 20$ and $30$ Trotter iterations per interval of time between data samples. The reason for such large interval of time between data samples is that the population terms evolution is much slower than the coherence terms evolution. In the detection stage of the 2DES experiment, we considered $t_3 = t_1$ ($t_3 = 1.45t_1$) and $N_3 = D_3 = D_1$ leading to $\Delta\omega_{3} = \Delta\omega_{1}$ ($\Delta\omega_{3} \approx \Delta\omega_{1}$) for the SQSP (PQP) simulation. 

Per Eq.~\eqref{eq_HI_pc}, the light-system interaction for each implemented pulse can be written as $\hat{H}_{I,pc}(t_p = d\Delta t_p,\phi)=\sum_{m}\alpha_{m}(d\Delta t_p,\phi) \hat{X}_{m}$, where $d$ denotes the Trotter iteration up to $D_p$ Trotter iterations. We chose an approximately instant Dirac-delta pulse profile, with $\alpha_{1}(0,\phi) \approx -8\times 10^3$~$cm^{-1}$ and $\alpha_{2}(0,\phi) \approx 0.8 \alpha_{1}(0,\phi)$~$cm^{-1}$. The phases $\phi_{j}$ are cycled for each pulse $j \in \{1,2,3\}$ over the set $\{0,2\pi/3,4\pi/3\}$ using the $3\times 3\times 3$ pulse scheme, yielding $27$ combinations.

In the PQP simulation, the system-probe interaction coupling strength written in the site basis during the interval of time $t_3$ is $J^{(pr)}=J^{(pr)}_{1}=J^{(pr)}_{2}=J_{1,2}/10$, which we found to be low enough to minimize back-reaction from the probe to the system. Considering the condition for the choice of $t_3$ shown in Eq.~\eqref{eq_condition_t3_main_text}, we find that accurate results can be obtained when $118<t_3\lesssim 1667$~$fs$, following the previously mentioned simulation parameters, hence we are in the perturbative regime of system-probe interaction.

Lastly, we considered a noiseless probe qubit and a constant single-qubit dephasing noise rate $\gamma_{Z} = 4$~$cm^{-1}$ acting on the system qubits. We implement the constant dephasing rate across the quantum cicuit simulation by renormalizing the probability of dephasing error on each Trotter layer depending on the chosen Trotter time-step for that Trotter layer. More specifically, we consider noiseless implementations of Trotter layers and, before and after each of these we add single-qubit dephasing channels written as $\mathcal{E}_{Z}(\hat{\rho}_{S})=(1-p_{Z}/2)\hat{\rho}_{S} + (p_{Z}/2)\hat{Z}\hat{\rho}_{S}\hat{Z}$ applied to each system's qubit, where $p_{Z}=\gamma_{Z} \Delta t_{j}$ is the probability of single-qubit dephasing error with $\Delta t_{j}$ denoting the Trotter time-step in the time interval $t_j$ and $j\in \{1,2,3\}$. The $\hat{Z}$ error probability introduced in each Trotter layer was $p_{Z} \approx 10^{-3}$. The double application of the noise channel in each Trotter layer is consistent with the second-order Trotter-Suzuki formula employed to evolve the quantum system.

In the PQP simulation, we compute the Fourier transform over $t_1$ of the measurement outcomes $\langle\hat{Y}_{pr}(t_1,t_2,\omega_{pr})\rangle_{t_3}$ and $\langle \hat{X}_{pr}(t_1,t_2,\omega_{pr})\rangle_{t_3}$ in the infinite-shot regime and store the measured (single-qubit) expectation values that allow us later to calculate $|\langle\hat{Y}_{pr}(\omega_1,t_2,\omega_{pr})\rangle_{t_3} -i \langle\hat{X}_{pr}(\omega_1,t_2,\omega_{pr})\rangle_{t_3}|$. We then normalize the value of each peak in the 2D spectra to the maximum of the results shown in Figs.~\ref{fig:results_main_text}(a,b). 

In the SQSP simulations, we perform simulations of 2DES experiments with phase-cycling (in the infinite-shot regime) following Sec.~\ref{sec: standard} and store the measured ($n_{qub}$) expectation values that allow us to later calculate the detection frequency $\omega_3$ line results as $|\langle \hat{F}(\omega_1,t_2,\omega_{3})\rangle|$. We then normalize the values with respect to the maximum of the results shown in Figs.~\ref{fig:results_main_text}(c,d).
To calculate the fluorescence observable in Eq.~\eqref{eq_F_obs} we chose $\Gamma_{1} = 1$ and $\Gamma_{2} = 2$. Lastly, in the SQSP simulation results, we changed the sign of the $\omega_1$ frequencies in the 2D spectra \cite{tan2008theory} for ease of results' comparison. 

\section{Peak dynamics}\label{sec:peak_evolution}
In Fig.~\ref{fig:peak_evol}, we plot the dynamics of several peaks of the 2D spectra with the SQSP and PQP for several choices of $J^{(pr)}t_3$. Overall, we observe that the PQP simulation can yield similar quantitative results as SQSP. Namely, in Figs.~\ref{fig:peak_evol}(a,d), we observe that the results with $J^{(pr)}t_3 = 0.15$ have the smallest error relatively to the corresponding peak dynamics of the SQSP simulation. On the other hand, in Fig.~\ref{fig:peak_evol}(b), the results with $J^{(pr)}t_3 = 0.22$ show that this choice has the decaying oscillation mean and amplitude that better matches the one of SQSP. In Fig.~\ref{fig:peak_evol}(c), it is shown that the results with $J^{(pr)}t_3 = 0.17$ have the smallest error relatively to the corresponding peak dynamics of SQSP. Therefore, we find that to obtain the smallest error of a PQP simulation relatively the SQSP and satisfy the condition of PQP applicability in Eq.~\eqref{eq_condition_t3_main_text}, the choice of $J^{(pr)}t_3$ should be in the interval $J^{(pr)}t_3 \in [0.15,0.22]$, with the results for $J^{(pr)}t_3 = 0.22$ providing the closest match to the SQSP in our simulations, and these are shown in Fig.~\ref{fig:results_main_text}. We also note that the choice of $J^{(pr)}t_3 \in [0.15,0.22]$ captures qualitatively well the oscillation dynamics of the peaks, an important feature to determine how coherent is the dynamics of the energy transport in photosynthetic systems \cite{engel2007evidence,panitchayangkoon2010long,thyrhaug2018identification}.

\section{Resource estimation details}\label{sec:resource_details}

The pulses implemented in the experiment \cite{thyrhaug2018identification} were centered at wavelength $\lambda = 805$~$nm$ and had nonzero amplitude in a time window of $14$~$fs$, which can be simulated by the SQSP and PQP following the procedure outlined in Sec.~\ref{sec: impl_phase_cycling}. For instance, we estimate that the implementation of the pulses with profile $E(t) = \exp{-2\ln(2)[(t-t_d)/\tau]^2}$, where $\tau = 5$~$fs$ and delay $t_d = 6.93$~$fs$ to center the pulse, require $D_p = 42$ Trotter layers for a pulse duration $t_p \approx 14$~$fs$. In the experiment of Ref.~\cite{thyrhaug2018identification}, the 2D spectra was obtained with frequency resolutions $\Delta \omega_1 = 36$~$cm^{-1}$, $\Delta \omega_3 = 40$~$cm^{-1}$ and during $t_1$ and $t_3$ ($t_2 = 1800$~$fs$), the data points were collected in intervals of $1.8$~$fs$ ($20$~$fs$). The maximum time of an experiment run, or equivalently circuit run in our protocol, was $3.56$~$ps$.

To estimate the number of measurements as a function of detection frequency resolution required for the SQSP simulation with the given frequency resolutions, we find that $N_1 = 515$, $N_2 = 90$ and $N_3 = 1/(\Delta \omega_3 \Delta t_3)$ data points must be collected on the time intervals $t_1$, $t_2$ and $t_3$, respectively, according to Eq.~\eqref{eq_meas_standard}. Hence, a minimum number of maximum Trotter layers in the quantum circuit of $D_1 \geq N_1$, $D_2 \geq N_2$, $D_3 \geq N_3$ and $D_p \geq 42$ for each time interval $t_1$, $t_1$ and $t_3$ and $t_p$, respectively, must be simulated. Using $D^{(SQSP)}_{max} = D_1 + D_2 + D_3 + 4D_p$ leads, in the case of $\Delta \omega_3 = 40$~$cm^{-1}$, to a maximum of $D^{(SQSP)}_{max} \geq 1237$ Trotter layers in the SQSP simulation. The number of system Hamiltonian queries, or equivalently, simulation runtime is given for the SQSP simulation as $Q_{SQSP} = S_{SQSP} Q_1 Q_2 Q_3$, where $Q_j = D_j (D_j+1)/2$ for $t_j$ with $j \in \{1,2,3\}$. We have considered $S_{SQSP}/S_{PQP}=(n_{qub}-1)=7$ with $n_{qub}=8$ since the ratio of the number of shots between the SQSP and PQP scales with system's size (see Eq.~\eqref{eq_S_sqsp}; we did not include the accuracy $\varepsilon$ in the calculation). The number of measurements is given by Eq.~\eqref{eq_meas_standard}, as $M_{meas}^{(SQSP)} = 27 \times S_{SQSP} \times D_1 \times D_2 \times D_3$, where $S_{SQSP} = 7$.

In the PQP simulation, we considered a number of measurements as given in Eq.~\eqref{eq_Mmeas_pr} where the number of relevant detection frequency lines is $N_{freq}=2$ \cite{thyrhaug2018identification}. Hence, we have $M_{meas}^{(PQP)} = 27 \times S_{PQP} \times D_1 \times D_2 \times N_{freq}$, where $S_{PQP} = 1$. The maximum number of Trotter layers is $D^{(PQP)}_{max} \geq 1403$, where we considered $D^{(PQP)}_{max} = D_1 + D_2 + D'_3 + 3D_p$. Here, $D'_3 = \alpha_{c} D_3$ where $\alpha_{c}=1.45$ denotes the constant found in the simulations performed for Fig.~\ref{fig:results_main_text} and it is introduced here because of the hidden constant in the lower bound of Eq.~\eqref{eq_condition_t3_main_text}. To compute the simulation runtime, we calculate the number of queries to the system Hamiltonian as $Q_{PQP} = Q_1 Q_2 Q'_3 N_{freq} $, where $Q'_3 = \alpha_{PQP} D'_3$ and $\alpha_{PQP} = 2.5$ is the assumed ratio of implementation time between a Trotter layer $\hat{T}_1^{(SP)}$ with $8$ qubits (requiring $\sim 10$ circuit layers) and $2$ qubits (requiring $\sim 4$ circuit layers).

\bibliography{apssamp}% Produces the bibliography via BibTeX.

\end{document}